\documentclass[11pt,letterpaper]{JHEP3}

  \usepackage{latexsym,bm,amsmath,amssymb,amsfonts}
  \usepackage{epsfig,graphics,graphicx}
  \usepackage{slashed}
  \usepackage[latin1]{inputenc}
  \usepackage{mathrsfs}
\usepackage{mathcomp}
\usepackage{yfonts}

  \long\def\comment#1{ }

  \def\nc {N_{\rm c}}
  \def\nf {N_{\rm f}}

\newcommand{\labell}[1]{\label{#1}}
\def\sac{\, , \,\,\,\,\,}

   \newcommand{\z}{\zeta}
  \newcommand{\eg}{{\it e.g.}}
  \newcommand{\ie}{{\it i.e.,}\ }
  \newcommand{\eqnum}[1]{Eq.~\eqref{#1}}

  \newcommand{\del}{\partial}

  \newcommand{\mcal}{\mathcal}
  \newcommand{\rme}{{\rm e}}
  \newcommand{\rmd}{{\rm d}}   %ELS%

  \newcommand{\nn}{\nonumber\\}
   
  \newcommand{\order}[1]{\mcal{O}{(#1)}}

  \newcommand{\beq}{\begin{eqnarray}}
  \newcommand{\eeq}{\end{eqnarray}}
  
 \def\simge{\mathrel{%
   \rlap{\raise 0.511ex \hbox{$>$}}{\lower 0.511ex \hbox{$\sim$}}}}
\def\simle{\mathrel{
   \rlap{\raise 0.511ex \hbox{$<$}}{\lower 0.511ex \hbox{$\sim$}}}}

\vspace{1.5cm}

\title{\rm \LARGE \bf Light--like mesons and deep inelastic scattering in
finite--temperature AdS/CFT with flavor}

\author{E. Iancu\\Institut de Physique Th\'eorique,
CEA Saclay, CNRS (URA 2306),
 F-91191 Gif-sur-Yvette, France\\
  E-mail:        \email{edmond.iancu@cea.fr}}
\author{A. H.~Mueller\\Department of Physics, Columbia University, New York, NY
10027, USA\\
        E-mail: \email{amh@phys.columbia.edu}}

\abstract{We use the holographic dual of a finite--temperature,
strongly--coupled, gauge theory with a small number of
flavors of massive fundamental
quarks to study meson excitations and deep inelastic scattering (DIS)
in the low--temperature phase, where the mesons are stable.
We show that a high--energy flavor current with nearly light--like
kinematics disappears into the plasma by resonantly producing vector mesons
in highly excited states. This mechanism generates the same DIS structure
functions as in the high temperature phase, where mesons are unstable
and the current disappears through medium--induced parton branching.
To establish this picture, we derive analytic results for the meson
spectrum, which are exact in the case of light--like mesons and which
corroborate and complete previous, mostly numerical, studies in the
literature. We find that the meson levels are very finely spaced near
the light--cone, so that the current can always decay, without a
fine--tuning of its kinematics.
}

\begin{document}

\section{Introduction}
\setcounter{equation}{0}

Motivated by some experimental results at RHIC, which suggest that the
deconfined matter produced in the intermediate stages of a
ultrarelativistic nucleus--nucleus collision might be strongly
interacting, there is currently a large interest towards understanding
the properties of strongly coupled field theories at finite temperature
within the framework of the AdS/CFT correspondence (see the review papers
\cite{MaldPhysRept,Son:2007vk,Iancu:2008sp,Erdmenger:2007cm,Gubser:2009sn,Rangamani:2009xk}
for details and more references). A substantial part of this effort has
been concentrated on studying the response of such a plasma to energetic,
`hard', probes, so like heavy quarks
\cite{Herzog:2006gh,Gubser:2006bz,Liu:2006ug,CasalderreySolana:2006rq,Gubser:2006nz,CasalderreySolana:2007qw,Gubser:2007xz,Chesler:2007sv,Dominguez:2008vd,Giecold:2009cg},
mesons (or quark--antiquark pairs)
\cite{Peeters:2006iu,Liu:2006nn,Chernicoff:2006hi,Caceres:2006ta,Argyres:2006vs,Ejaz:2007hg,mateos,Hoyos:2006gb,starinets},
or photons
\cite{HIM2,HIM3,CaronHuot:2006te,Mateos:2007yp,CasalderreySolana:2008ne,Bayona:2009qe},
in an attempt to elucidate some intriguing RHIC data, so like the
unexpectedly large `jet quenching', or to provide alternative signatures
of a strongly--coupled matter.

In particular, AdS/CFT calculations of deep inelastic scattering (DIS)
\cite{Polchinski:2002jw,HIM1,
HIM2,HIM3,Bayona:2009qe,BallonBayona:2007rs,Albacete:2008ze,Mueller:2008bt,Avsar:2009xf,Dominguez:2009cm}
have given access to the structure of strongly coupled matter at high
energy and for small space--time separations and thus revealed an
interesting picture, which is quite different from the corresponding
picture in a gauge theory at weak coupling. One has thus found that there
are no point--like `partons' at strong coupling, that is, no constituents
carrying a sizeable fraction $x\sim\order{1}$ of the total longitudinal
momentum of a `hadron' \cite{Polchinski:2002jw,HIM1} or plasma
\cite{HIM2,HIM3} at high energy. This has been interpreted as the result
of a very efficient branching process through which all partons have
fallen at the smallest values of $x$ which are consistent with energy
conservation. This interpretation is consistent with arguments based on
the operator product expansion at strong coupling
\cite{Polchinski:2002jw,Iancu:2009zz} and is further supported by the
fact that the DIS structure functions were found to be large at small
values of $x\ll 1$, where they admit a natural interpretation in terms of
partons \cite{HIM1,HIM2,Iancu:2008sp}. The central scale in this picture
is the saturation momentum $Q_s(x)$, which defines the borderline between
the large--$x$ and large virtuality ($Q>Q_s(x)$) domain which is void of
partons, and the small--$x$, small virtuality ($Q\lesssim Q_s(x)$) domain
where parton exist, with occupation numbers of order 1 (a situation
somewhat reminiscent of, but more extreme than, parton saturation in QCD
at weak coupling \cite{CGCreviews}). This scale $Q_s$ plays an essential
role also for other high energy processes, of direct relevance to heavy
ion collisions, like the energy loss and the momentum broadening of a
heavy quark \cite{Dominguez:2008vd,Giecold:2009cg}. It can furthermore be
related to the dissociation length for a large, semiclassical, `meson'
\cite{Liu:2006nn,Chernicoff:2006hi,Caceres:2006ta,Argyres:2006vs}.

So far, most studies of DIS at finite temperature and strong coupling
were concerned with the ${\mathcal N}=4$ supersymmetric Yang--Mills (SYM)
theory, where the role of the virtual photon is played by the ${\cal
R}$--current --- a conserved current associated with a U(1) global
symmetry which couples to massless fields in the adjoint representation
of the color group SU$(\nc)$. Very recently, the same problem has been
addressed \cite{Bayona:2009qe} within the context of a ${\mathcal N}=2$
supersymmetric plasma obtained by adding $\nf$ hypermultiplets of
(generally massive) fundamental fields to the ${\mathcal N}=4$ SYM
plasma, in the `probe' limit where $\nf\ll \nc$. In that case, the
`photon' is a flavor current which couples to a pair of fields (fermions
and scalars) in the fundamental representation of SU$(\nc)$, that we
shall refer to as `quarks'. In order to describe the results in
Ref.~\cite{Bayona:2009qe} and also our subsequent results in this paper,
it is useful to briefly recall the structure of the holographic dual of
the ${\mathcal N}=2$ plasma at strong coupling $\lambda\equiv g^2\nc\gg
1$, also known as the `D3/D7 model'
\cite{Karch:2002sh,Babington:2003vm,Kruczenski:2003be,Erdmenger:2007cm}
(see also Sect.~\ref{sec:D3D7} below).

The supergravity fields (in particular, the Abelian gauge field dual to
the flavor current) live in the worldvolume of one of the $\nf$
D7--branes that have been inserted in the AdS$_5\times S^5$ Schwarzschild
background geometry dual to the ${\mathcal N}=4$ SYM plasma. For the case
where the fundamental fields are massive, the D7--branes are separated in
the radial direction from the $\nc$ D3--branes located at the `center' of
AdS$_5\times S^5$. The distance between the two systems of brane then
fixes the `bare' mass of the fundamental `quarks', represented by
Nambu--Goto strings stretching from a D7--brane to a D3--brane.
Flavorless `mesons', or quark--antiquark bound states, can be described
either as strings with both endpoints on a D7--brane, or (at least for
sufficiently small meson masses and spins) as normal modes of the
supergravity fields propagating in the worldvolume of the D7--brane. In
this paper, we shall adapt the second point of view, that of the normal
modes.

For zero or sufficiently low temperatures, the mesons are strongly bound
\cite{Kruczenski:2003be,mateos}: their binding energy almost compensates
the large quark masses (proportional to the string tension), so that the
meson masses remain finite in the strong coupling limit
$\lambda\to\infty$ (the `supergravity approximation'), to which we shall
restrict ourselves. In that limit, the quarks become infinitely massive
and then the mesons are stable: they form an infinite, discrete, tower of
(scalar and vector) modes, distinguished by their quantum numbers. A
remarkable property of the meson spectrum that will play an essential
role in what follows is that, at finite temperature, the dispersion
relation for a given mode changes virtuality, from time--like to
space--like, with increasing momentum \cite{mateos}. This is illustrated
in Fig.~\ref{fig:dispersion}. In particular, there exists an intermediate
value of the momentum at which the mode becomes light--like.

\FIGURE[t]{
\centerline{%\hspace*{-3.cm}
\includegraphics[width=0.65\textwidth]{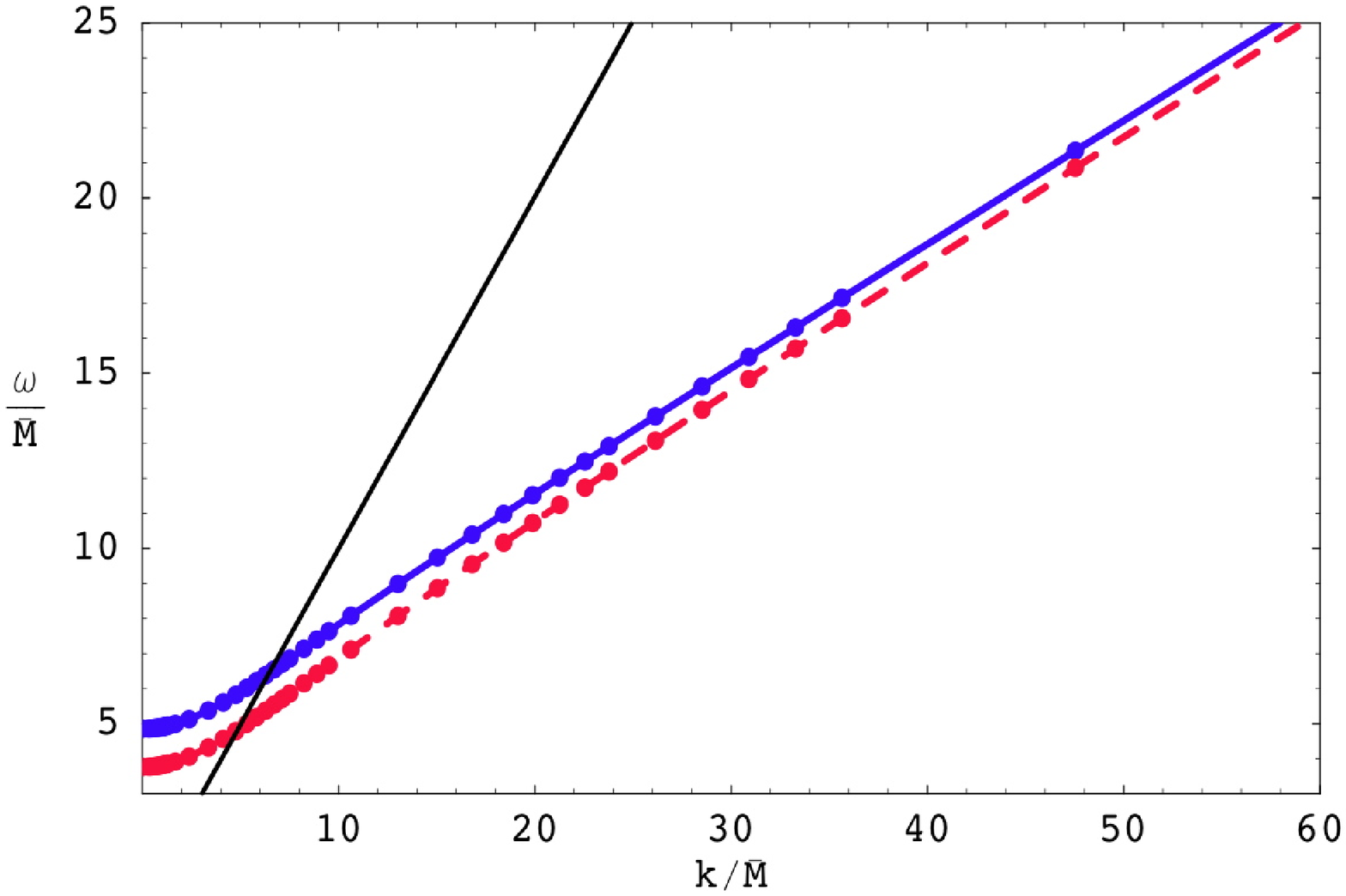} }
\caption{\sl Dispersion relation
$\omega(k)$ for the lightest spin--zero mesons in the low temperature
phase, or Minkowski embedding \cite{mateos}. The solid blue curve
corresponds to a pseudo--scalar meson, whereas the red dashed curve
corresponds to a scalar meson. The solid black line corresponds to
$\omega = k$. \label{fig:dispersion}}}

This situation persists for sufficiently low temperatures, so long as the
D7--brane, although deformed by the attraction exerted by the black hole,
remains separated from the latter. But with increasing temperature, one
finds a first--order phase transition (from the `Minkowski embedding' to
the `black hole embedding') at some critical temperature $T_c$, at which
the tip of the D7--brane suddenly jumps into the black hole (BH) horizon
\cite{Mateos:2006nu,mateos}. For $T\ge T_c$, all the mesons `melt' :
their dispersion relations acquire large imaginary parts (comparable to
their real parts), showing that the bound states are now highly unstable
\cite{Hoyos:2006gb,starinets}.

One should stress that this peculiar `meson melting' phase transition is
specific to this model and has no analog in QCD. The corresponding
situation in QCD is not yet fully clear\footnote{For instance, potential
models using lattice QCD input predict that all charmonium states and the
excited bottomonium states dissolve in the QGP juste above $T_c$ (see the
recent review \cite{Mocsy:2009ca}). But the most refined, recent, lattice
calculations cannot exclude the survival of charmonium states up to
$T=1.5T_c$ \cite{Ding:2009ie}.}, but on physical grounds (given that the
deconfinement `phase transition' is truly a cross--over) one would expect
the mesons to gradually melt when increasing $T$ above $T_c$, according
to their sizes. The smallest mesons, those built with the heavy quarks,
may survive in the temperature range pertinent to RHIC or LHC. Assuming
the quark--gluon plasma to be effectively strongly coupled within this
range, one could use the `low--temperature' phase of this model to get
some insight into the properties of the still surviving, very small and
very heavy, mesons, and the `high--temperature' phase for the larger,
lighter, mesons that have already melt.

Returning to the problem of the DIS for the flavor current, it is quite
clear that the situation is very different in the high--temperature phase
($T\ge T_c$) as compared to low--temperature one ($T< T_c$). At $T
> T_c$, the problem is conceptually the same as for the ${\cal
R}$--current in the ${\mathcal N}=4$ plasma \cite{HIM2,HIM3}. The
space--like current fluctuates into a system of virtual partons
--- for the flavor current, this system involves a pair of fundamental
fields together with arbitrary many ${\mathcal N}=4$ quanta
--- whose subsequent evolution depends upon the kinematics. If the energy
$\omega$ of the current is low enough (for a fixed virtuality $Q$), the
partonic fluctuation closes up again and essentially nothing happens: the
space--like current is stable, so like in the vacuum, due to
energy--momentum conservation. (We ignore here tunnel effects at finite
temperature, which are exponentially small \cite{HIM2,Mueller:2008bt}.)
But if the energy is sufficiently high, such that $\omega\gtrsim
Q^3/T^2$, the virtual partons live long enough to feel the interactions
with the plasma (in the dual gravity problem: the attraction exerted by
the black hole) and under the influence of these interactions they keep
branching until they disappear into the plasma. In the dual gravity
theory, this medium--induced branching is seen as the fall of the dual
gauge field into the BH horizon. Since this process is fully driven by
interactions with the ${\mathcal N}=4$ plasma, the saturation momentum
and also the DIS structure functions at small $x$ should be the same as
for the ${\cal R}$--current, up to a global factor which counts the
number of degrees of freedom to which couples the current ($\nc^2$ for
the ${\cal R}$--current and, respectively, $\nc\nf$ for the flavor one).
This is indeed what was found in Ref.~\cite{Bayona:2009qe} for the
high--temperature phase.

But at lower temperatures $T< T_c$, the situation turns out to be more
subtle. On the supergravity side, it is {\em a priori} clear, from
geometrical considerations, that the gauge field dual to the flavor
current cannot fall into the black hole, for any energy: indeed, the
support of this field is restricted to the worldvolume of the D7--brane,
which is now separated from the BH horizon along the radial direction.
This reasoning led the authors in Ref.~\cite{Bayona:2009qe} to conclude
that DIS should not be possible in this case, however high is the energy.
It is understood here that the high--energy limit does not commute with
the large--$\lambda$ limit: the energy must remain small not only as
compared to the quark mass $m_{\rm q}\sim \sqrt{\lambda} M_{\rm gap}$,
but also as compared to the mass of the lowest string excitations $m_{\rm
string}\sim {\lambda}^{1/4} M_{\rm gap}$ \cite{Kruczenski:2003be}, which
are not described by the supergravity fields. Here, $M_{\rm gap}$ is the
lowest meson mass at zero temperature, and is independent of $\lambda$
(see Sect.~\ref{sec:D3D7} below).

Our main, new, observation is that, although it cannot fall into the BH,
the flavor current can nevertheless disappear into the plasma by
resonantly producing space--like vector mesons which, as already
mentioned, are indeed supported by the plasma. (For a time--like current
in the vacuum, the resonant production of mesons has been discussed in
App. B of Ref.~\cite{starinets}.) For this process to be possible, the
kinematics of the current should match with the dispersion relations for
the vector mesons. For this interaction to qualify as `deep inelastic
scattering', the associated structure functions --- determined by the
coupling of the current to the mesons and computed as the imaginary part
of the current--current correlator --- must be non--zero in a continuous
domain of the phase--space, and not only at discrete values of the
energy. In this paper, we shall demonstrate that these conditions are
indeed satisfied at sufficiently high energy. Our final result is that
the structure functions for flavor DIS are exactly the same in this low
temperature--phase as in the high--temperature phase, although the
respective physical pictures are quite different. This result is in fact
natural, as we shall later explain.

To develop our arguments, we shall perform a detailed study of the meson
excitations in the high--energy, space--like, kinematics relevant for DIS
off a strongly coupled plasma; that is, $\omega\gg Q\gg T$, with
$Q^2\equiv k^2-\omega^2>0$. We shall focus on vector mesons with
transverse polarizations, which provide the dominant contribution to DIS
at high energy \cite{HIM2}, but we expect similar results to apply for
other types of excitations (longitudinal vector mesons, scalar and
pseudoscalar ones). Also, for technical reasons, we shall limit ourselves
to the case of very heavy mesons, or very low temperature, $M_{\rm
gap}\gg T$, which however captures all the salient features of the
general situation.

Concerning the kinematics, we shall find that a space--like current can
excite mesons only for high enough energies and relatively small
virtualities, such that the current and the mesons are nearly
light--like. This is so because a current with large space--like
virtuality encounters a potential barrier near the Minkowski boundary
(associated with energy--momentum conservation) and thus cannot penetrate
in the inner region of the D7--brane, where mesons could be created.
However, for high enough energy $\omega\gtrsim Q^3/T^2$,  this barrier is
overcome by the gravitational attraction due to the black hole (\ie by
the mechanical work done by the plasma \cite{HIM2,HIM3}), and then the
current can penetrate inside the bulk and thus excite mesons. The
corresponding kinematics being nearly light--like, $\omega\simeq k$, we
shall focus our attention on the respective region of the meson
dispersion relation in Fig.~\ref{fig:dispersion}, but we shall provide
analytic approximations also for the other regions (in
Sect.~\ref{sec:Mesons}). Our results are as follows.

For the strictly light--like mesons ($\omega=k$), we shall construct in
Sect.~\ref{sec:resDIS} exact, analytic, solutions for the spectrum and
the wavefunctions, which take particularly simple forms for large quantum
numbers $n\gg 1$. We shall thus find an infinite tower of equally spaced
levels, with high energies and a large level spacing: $\omega_n\simeq
n\Delta\omega$ where $\Delta\omega\sim T(M_{\rm gap}/ T)^3\gg M_{\rm
gap}$. (For comparison, at zero momentum, the energy of the mode $n$ is
$\omega_n(k=0)\sim nM_{\rm gap}$.) Similarly, for the gauge field dual to
a light--like flavor current, we shall find exact `non--normalizable'
solutions, from which we shall compute the retarded current--current
correlator in the high energy limit $\omega\gg T(M_{\rm gap}/ T)^3$. As
expected, this propagator exhibits poles at the energies of the
light--like mesons, so its imaginary part is an infinite sum over
delta--like resonances. The coefficient of each delta--function
represents the probability for the resonant production of a meson by a
current whose energy is exactly $\omega_n$. Conversely, they also
describe the rate for the decay of a vector meson into an on--shell
photon, a mechanism recently proposed as a possible signal of strong
coupling behaviour in heavy ion collisions
\cite{CasalderreySolana:2008ne}.

The resonant production of mesons remains possible also for slightly
space--like kinematics, and, of course, for any time--like kinematics,
but this is perhaps not the most interesting physical situation, as it
requires the energy of the current to be finely tuned to that of a meson
mode. Given the large level spacing $\Delta\omega$ indicated above, it
looks at a first sight unlikely that a small uncertainty $\delta\omega\ll
\omega$ in the energy of the current --- as inherent in any scattering
experiment (even a {\em Gedanken} one !), where the `photon' is not a
plane wave, but a wave packet --- could help reducing the need for the
fine--tuning. If that was true, it would mean that for the whole
high--energy region in phase--space, except for a set of zero measure (as
defined by the dispersion relations for the meson modes), the current
survives in the plasma for arbitrarily long time. However, as we shall
argue now, that conclusion would be a bit naive, as it underestimates the
consequences of a small fluctuation in the energy, or the virtuality, of
the current for the problem at hand.

The main point is that the energy uncertainty $\delta\omega$ should not
be compared to the (relatively large) level spacing $\Delta\omega$
between two successive resonances, but rather to the change in energy
which is necessary to cross from one meson level to another {\em at a
fixed value $k$ of the momentum}. Indeed, one should not forget that the
dispersion relation in the relevant kinematics is nearly light--like,
that is, $\omega_n\simeq k_n$ for the $n$th mode. Hence, when increasing
the energy by $\Delta\omega$ to move from one level to a neighboring one,
one is simultaneously increasing the momentum $k$ by the same, large,
amount --- one moves along the light--cone. But in a scattering problem,
the momentum $k$ of the current is fixed and its energy $\omega$ has
generally an uncertainty $\delta\omega$, related to the fact that the
source producing the current has been acting over a finite period of time
$\delta t$: $\delta\omega\sim 1/\delta t$. Before we discuss this time
$\delta t$, let us make the crucial observation that the meson levels are
very finely spaced in energy when probed at a fixed value of $k$. This is
a general feature of the high--energy kinematics, which is further
amplified by the peculiar shape of the meson dispersion relation near the
light--cone.

\FIGURE[t]{
\centerline{%\hspace*{-3.cm}
\includegraphics[width=0.7\textwidth]{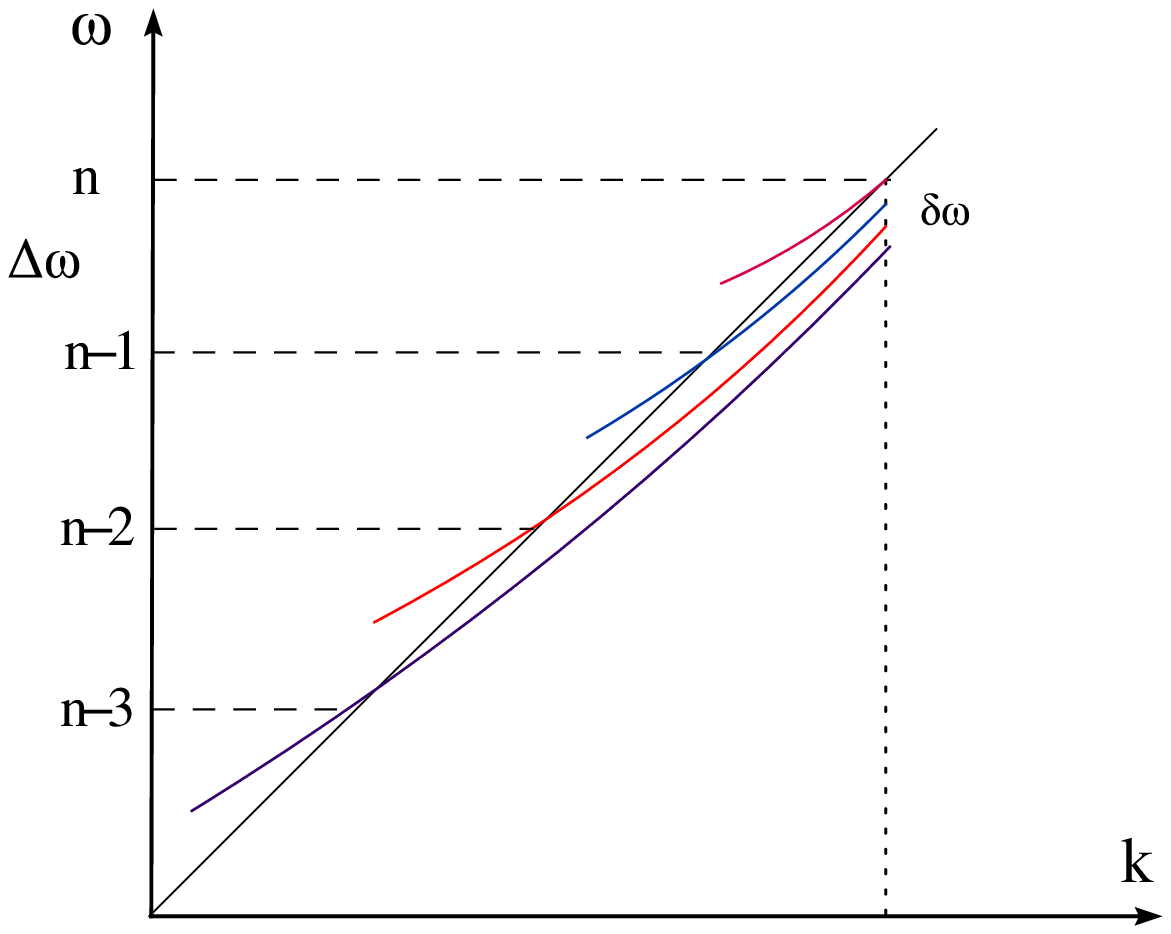} }
\caption{\sl Qualitative illustration of the meson dispersion relation
$\omega(k)$ in the vicinity of the light--cone. Four successive modes are
shown. Although widely spaced along the light--cone, the modes are close
to each other at any given value of $k$. Thus, in order to jump from one
mode to a neighboring one, one needs a relatively large energy jump
$\Delta\omega$ at fixed virtuality $Q$, but only a small energy jump
$\delta\omega$ at fixed momentum $k$. \label{fig:dispLL}}}

As a simpler example, recall first the situation at zero temperature
\cite{Kruczenski:2003be}, where the meson dispersion relation reads,
schematically, $\omega_n(k)=(k^2+n^2M_{\rm gap}^2)^{1/2} \simeq k +
n^2M_{\rm gap}^2/2k$, with the approximate equality holding when $k\gg n
M_{\rm gap}$. Hence the energy jump $\delta\omega_n(k)\equiv
\omega_{n+1}(k)-\omega_n(k)$ needed to cross from one mode to another at
fixed $k$ is $\delta\omega_n(k)\simeq nM_{\rm gap}^2/k$ and becomes
smaller and smaller when increasing $k$. As anticipated, the modes are
very finely spaced at large $k$. Returning to the finite--$T$ case of
interest, it turns out that the respective dispersion relation is even
more sensitive to small variations in the virtuality of the mode near the
light--cone. Specifically, we shall find in Sects.~\ref{sec:LL} and
\ref{sec:TIME} that the level spacing at fixed $k$ defined as above
scales with $k$ as $\delta\omega_n(k)\sim T(T/k)^{1/3}$ when $k\simeq k_n
= n\Delta\omega$ (see Fig.~\ref{fig:dispLL}). As anticipated, this is
considerably smaller than the energy spacing $\Delta\omega\sim T(M_{\rm
gap}/ T)^3$ at fixed virtuality : indeed, $\Delta\omega/\delta\omega_n
\sim n^{1/3}(M_{\rm gap}/T)^4 \gg 1$.

To understand the typical energy uncertainty $\delta\omega$ of the
current, one needs an estimate for its interaction time in the plasma
$t_{\rm int}$. Indeed, the source producing the current should act over a
comparatively shorter time $\delta t\lesssim t_{\rm int}$ in order for
the subsequent dynamics to be observable. Via a time--dependent analysis
of the dynamics of the dual gauge field in Sect.~\ref{sec:TIME}, we shall
find that $t_{\rm int}$ is controlled by the progression of the gauge
field within the D7--brane, which yields $t_{\rm int}\sim (k/T)^{1/3}/T$
(similarly to the ${\cal R}$--current \cite{HIM3}). This estimate implies
a lower limit $\delta\omega\gtrsim 1/t_{\rm int}$ on the energy
uncertainty of the current which is of the order of the level spacing
$\delta\omega_n$ indicated above. This justifies performing an average
over neighboring levels in the calculation of the imaginary part of the
current--current correlator. This averaging smears out the meson
resonances and produces our main result in this paper, \eqnum{ImPi}. As
anticipated, this result is identical to the DIS structure functions in
the high--temperature phase, which shows that the current is completely
absorbed by the plasma in both cases.

The analysis in Sect.~\ref{sec:TIME} also allows us to deduce a
space--time picture for the nearly light--like mesons in the
semiclassical regime at large quantum numbers $n\gg 1$, where the notion
of a classical orbit makes sense. We thus find that the period for one
orbit is $\Delta t_n\sim (\omega_n/T)^{1/3}/T$, where we recall that the
energy of the bound state is $\omega_n= n\Delta\omega$. Furthermore, we
find that the meson spends the major part of this time far away from the
tip of the D7--brane, at relatively large radial distances $\sim
\omega_n^{1/3}$. This is so because its orbital velocity is much higher
near the tip than at larger radial distances. It is finally interesting
to notice that, in this light--like kinematics, the period $\Delta t_n$
of the bound state has the same parametric dependence upon its energy as
the interaction time $t_{\rm int}$ of the current, and similarly for the
typical radial location of the meson versus the saturation momentum
$Q_s(k)\sim k^{1/3}$ for the current.

\section{Mesons in the D3/D7 brane model at finite temperature}
\setcounter{equation}{0} \label{sec:D3D7}

%\subsection{The geometry of the model}

According to the AdS/CFT correspondance
\cite{Maldacena:1997re,Gubser:1998bc,Witten:1998qj}, the
four--dimensional ${\cal N}=4$ super--Yang--Mills (SYM) gauge theory with
`color' gauge group SU$(\nc)$ is dual to a type IIB string theory living
in the ten--dimensional curved space--time AdS$_5\times S^5$, which
describes the decoupling limit of $\nc$ black D3--branes. By further
adding a black brane to this geometry, one obtains the holographic dual
of the finite--temperature, plasma, phase of the ${\cal N}=4$ SYM
\cite{Witten:1998zw}. The ensuing metric reads (see, \eg,
\cite{MaldPhysRept})
\beq
\rmd s^2 = \frac{u^2}{L^2} \left( -f(u)\rmd t^2 + \rmd{\bm x}^2\right) +
\frac{L^2}{u^2} \left( \frac{\rmd u^2}{f(u)} +u^2 \rmd\Omega_5^2\right)\,
,  \label{BH}
\eeq
where $f(u) = 1-u_0^4/u^4$, with $u_0=\pi L^2 T$ the radial position of
the black hole horizon and $T$ the common temperature of the ${\cal N}=4$
SYM plasma and of the black hole. The curvature radius $L$ is defined in
terms of the string coupling constant $g_s$ and the string length scale
$\ell_s$ via $L^4 = 4\pi g_s \nc \, \ell_s^4$.  The holographic
dictionary relates the gauge and string theory coupling constants as $g^2
= 4 \pi g_s$. In the ``strong coupling limit'' of the gauge theory,
defined as $\nc\rightarrow \infty$, $\lambda\equiv g^2 \nc\rightarrow
\infty$, with $g$ fixed and small ($g\ll 1$), the string theory reduces
to classical supergravity theory in the AdS$_5\times S^5$ Schwarzschild
geometry with metric \eqref{BH}.

All fields in the ${\cal N}=4$ SYM theory transform in the adjoint
representation of SU$(\nc)$. Fields transforming in the fundamental
representation of the gauge group can be introduced in the gravity dual
by inserting a second set of D--branes in the supergravity background
\cite{Karch:2002sh,Erdmenger:2007cm}. In particular, we consider the
decoupling limit of the intersection of $\nc$ black D3--branes and $\nf$
D7--branes as described by the array:
\begin{equation}
\begin{array}{ccccccccccc}
   & 0 & 1 & 2 & 3 & 4& 5 & 6 & 7 & 8 & 9\\
{\rm D3} & \times & \times & \times & \times & & &  &  & & \\
{\rm D7} & \times & \times & \times & \times & \times
& \times & \times & \times &  &   \\
\end{array}
\label{array}
\end{equation}
where the first four dimensions (0, 1, 2, and 3) correspond to the
Minkowski coordinates $\{ t, x^i \}$ and the last six ones (from 5 to 9)
to the six--dimensional space with coordinates $\{u, \Omega_5\}$. The
dual field theory is now an ${\cal N}=2$ gauge theory consisting of the
original SYM theory coupled to $\nf$ fundamental hypermultiplets which
consists of two Weyl fermions and their superpartner, complex, scalars
(see \eg{} \cite{AlvarezGaume:1996mv}). For brevity, we shall globally
refer to these fundamental fields as `quarks'. In the limit where the
number of flavors is relatively small, $\nf \ll \nc$, the D7--branes may
be treated as probes in the black D3--brane geometry \eqref{BH}. That is,
the D7--branes are generally deformed by their gravitational interactions
with the D3--branes and the black hole, but one can neglect their back
reaction on the ambient geometry, \eqnum{BH}. The ensuing geometry is
dual to a ${\cal N}=2$ plasma at finite temperature in which the effects
of the fundamental degrees of freedom (say, on thermodynamical
quantities) represent only small corrections, of relative order $g^2 \nf
=\lambda (\nf/\nc)\ll 1$ (see \eg{} \cite{mateos}).

\FIGURE[t]{ \centerline{
\includegraphics[width=0.8\textwidth]{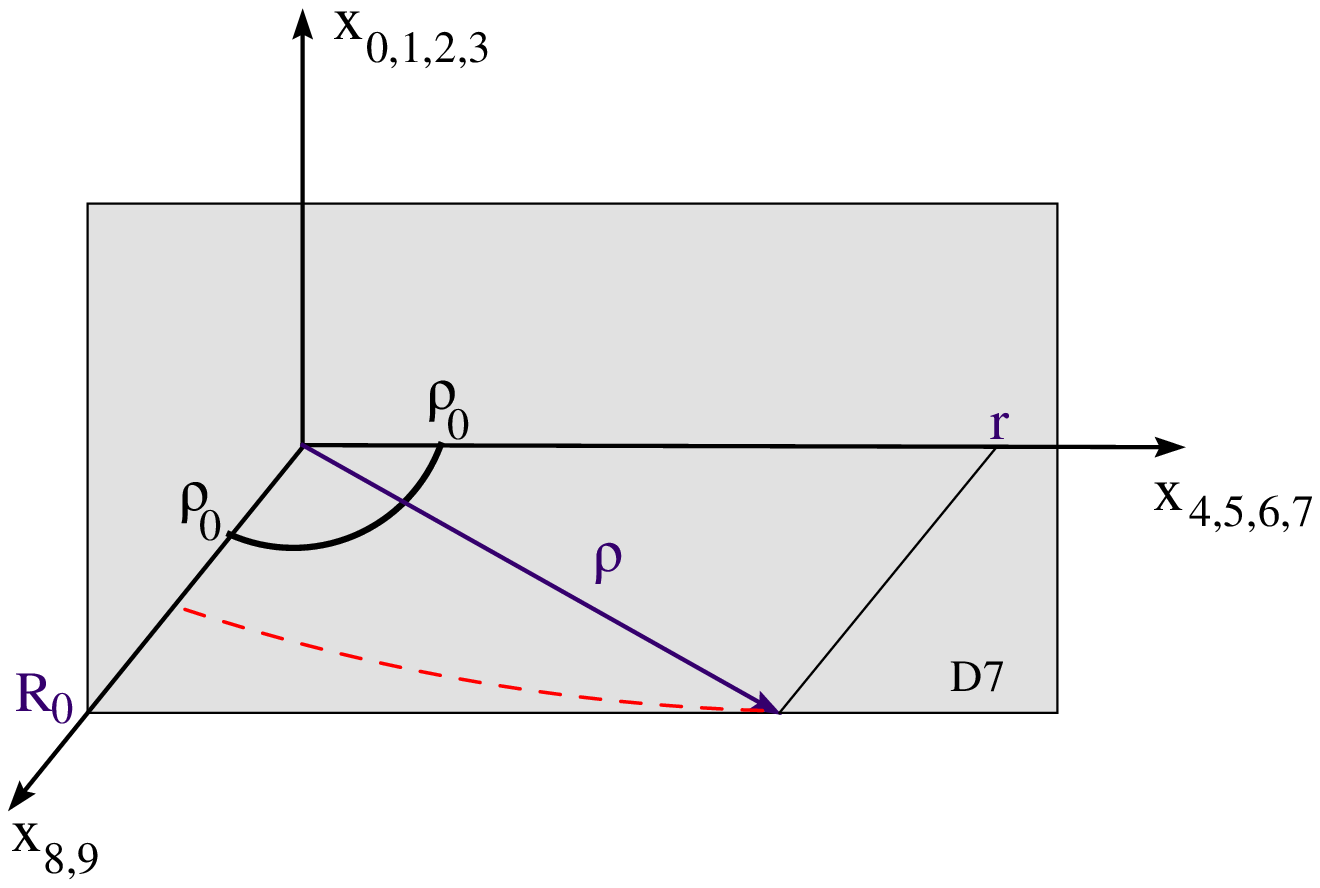}}
\caption{\sl Schematic representation of the geometry of the D3--D7
model at finite temperature. This drawing is adapted to the
low--temperature, or very heavy quark, situation, where $R_0\gg
\rho_0=1$. See the text for more explanations. \label{fig:D7brane}}}

Although both the D3--branes and the D7 ones fill the Minkowski space,
these two types of branes need not overlap with each other, as they can
be separated in the 89--directions, which are orthogonal to both of them.
When this happens, the conformal symmetry is explicitly broken already at
classical level\footnote{Quantum mechanically, the conformal symmetry is
broken by the D7--branes even when they overlap with the D3--branes, \ie
when $u_m=0$. But the $\beta$--function for the `t Hooft coupling
$\lambda=g^2\nc$ is of order $\nf/\nc$ and thus is suppressed in the
probe limit $\nf/\nc\to 0$.} and then the fundamental fields in the dual
gauge theory become massive: their `bare' mass is proportional to the
radial separation $u_m$ between the two sets of branes at zero
temperature. Indeed, a fundamental field is `dual' to an open string
connecting a D7--brane to a D3--brane, so its `bare' mass is equal to the
string length $u_m$ times the string tension:
 \beq\label{mq}
 m_{\rm q}\,=\,\frac{u_m}{2\pi\ell_s^2}\,=\,\sqrt{\lambda}\,
 \frac{u_m}{2\pi L^2}\,.\eeq
%Note that this mass is parametrically large, of $\order{\sqrt{\lambda}}$.

To render such geometrical considerations more suggestive, it is helpful
to perform some changes of coordinates \cite{Mateos:2006nu,mateos}.
First, we introduce a new, dimensionless, radial coordinate $\rho$,
related to the coordinate $u$ via\footnote{We notice that $\rho$ is
related to the Fefferman--Graham \cite{FG} radial coordinate $z$ via
$z/\sqrt{2}=L^2/(u_0\rho)$.}
\beq
(u_0 \rho)^2 = u^2 + \sqrt{u^4-u_0^4}\, . \label{change1}
\eeq
Note that that the BH horizon corresponds to $\rho_0=1$ and the Minkowski
boundary to $\rho\to\infty$, with $u_0 \rho\simeq\sqrt{2}u$ when $u\gg
u_0$ (\ie $\rho\gg 1$). Then the background metric \eqref{BH} becomes
\beq
\rmd s^2 = \frac{1}{2} \left(\frac{u_0 \rho}{L}\right)^2 \left[-{f^2\over
\tilde f}\,\rmd t^2 + \tilde{f} \rmd{\bm x}^2 \right] +
\frac{L^2}{\rho^2}\left[\rmd\rho^2 +\rho^2 \rmd\Omega_5^2  \right] \,,
\label{D3geom}
\eeq
where
\beq\label{ff}
f(\rho)= 1- \frac{1}{\rho^4} \sac \tilde{f}(\rho)=1+\frac{1}{\rho^4} \,.
\eeq
It is furthermore useful to adapt the metric on the five--sphere to the
D7--brane embedding. Since the D7--brane spans the 4567-directions, we
introduce spherical coordinates $\{r, \Omega_3\}$ in this space and $\{R,
\phi\}$ in the orthogonal 89--directions. Denoting by $\theta$ the angle
between these two spaces, we have (see also Fig.~\ref{fig:D7brane})
\beq
\rho^2=r^2+R^2 \sac r=\rho\cos\theta \sac R=\rho\sin\theta \,,
\labell{coord2}
\eeq
and therefore
\beq
\rmd\rho^2+\rho^2\rmd\Omega^2_5&=&
\rmd\rho^2+\rho^2\left(\rmd\theta^2+\cos^2\theta \, \rmd\Omega^2_3+
\sin^2\theta \, \rmd\phi^2\right) \,  \nn &=& \rmd r^2 +
r^2\rmd\Omega^2_3 + \rmd R^2+R^2\rmd\phi^2 \,. \label{change2}
\eeq
Note that, on the D7--brane, the Minkowski boundary lies at $r\to\infty$.

To specify the D7--brane (background) embedding, we require translational
symmetry in the 0123--space and rotational symmetry in the
4567--directions, and fix $\phi=0$. Then the embedding can be described
as the profile function $R=R_v(r)$. The subscript `$v$' on $R_v$ stays
for the `meson vacuum': the small fluctuations of the D7--brane around
its stationary geometry are dual to low--lying `mesons' in the boundary
gauge theory, \ie (colorless and flavorless) bound states which involve a
pair of fields from a fundamental hypermultiplet
--- say, a quark--antiquark pair. Such mesons are represented by strings
with both ends on the D7--branes and thus can be studied (at least for
small enough meson sizes and masses; see below) by examining the small
fluctuations of the worldvolume fields on the D7--branes. These include
the fluctuations $\delta\phi$ and $\delta R$ in the shape of the
D7--brane --- which give rise to pseudo--scalar and scalar mesons,
respectively ---, and also fluctuations of the worldvolume gauge fields,
which describe vector mesons. The `vacuum' profile function $R_v(r)$ and
the spectrum of the various type of fluctuations have been systematically
studied in the literature, via analytic methods in the zero--temperature
case \cite{Kruczenski:2003be}, and via mostly numerical methods at
non--zero temperature
\cite{Babington:2003vm,Mateos:2006nu,mateos,Hoyos:2006gb,starinets}. In
what follows, we shall collect the previous results which are relevant
for the present analysis, with a minimum of formul\ae.

The dynamics of the D7--brane is described the Dirac--Born--Infeld (DBI)
action\footnote{The full D7--brane action also involves a Wess--Zumino
term, but this plays a role only for those gauge field configurations
having non--trivial components along the three--sphere $S^3$ internal to
the D7--brane \cite{Kruczenski:2003be,starinets}. Such fields do not
enter the study of deep inelastic scattering and will not be considered
throughout this work.}. The profile function $R_v(r)$ for the `vacuum'
embedding is obtained by solving the equation of motion for $R(r)$ which
follows from this action. The meson spectrum is then obtained by solving
the linearized equations of motion (EOM) which follow after expanding the
DBI action to quadratic order in small fluctuations around the `vacuum'
embedding.

At zero temperature, one finds that the `vacuum' profile is trivial, \ie
independent of $r$~:
 \beq
 R_v(r)\,=\,R_0\,=\,\sqrt{2}\,\frac{u_m}{u_0}\qquad\mbox{
 (zero temperature)}\,,
 \eeq
where we recall that $u_m$ is the separation between the two types of
brane in the original radial coordinate $u$. (At $T=0$, \eqnum{change1}
reduces to $\rho\equiv \sqrt{2}(u/u_0)$ where $u_0$ is an arbitrary
reference scale, which drops out from the final results.) The EOM for the
small fluctuations have been solved exactly, in terms of hypergeometric
functions % (some solutions will be shown in the next section)
\cite{Kruczenski:2003be}. At zero temperature, both Lorentz symmetry and
supersymmetry are manifest. Accordingly, for a meson with four--momentum
$q^\mu=(\omega, 0, 0,k)$, the dispersion relation $\omega(k)$ involves
only the `invariant mass' combination $M^2\equiv \omega^2 - k^2$.
Besides, this relation depends upon two `quantum numbers': a `radial'
number $n=0,1,2,...$, which counts the number of zeroes of the
corresponding wavefunction in the interval $0 < r < \infty$, and an
`angular' number $\ell$, with $\ell=0,1,2,...$, which refers to rotations
along the $S^3$ component of the D7--brane. (In the dual gauge theory,
$\ell$ represents a charge under the internal symmetry group SO(4) which
is dual to rotations on $S^3$.) Supersymmetry together with the global
SO(4) symmetry imply additional degeneracies for the meson spectrum, as
discussed in \cite{Kruczenski:2003be}. Specifically,
Ref.~\cite{Kruczenski:2003be} found the following dispersion relation
 \beq\label{T0spectrum}
 M^2(n,\ell)\,\equiv\, \omega_{n\ell}^2(k) - k^2\,=\,
 \frac{u_m^2}{L^4}\,4(n+\ell+1)(n+\ell+2)\eeq
for both (pseudo)scalar and vector mesons. Note the presence of a mass
gap in the spectrum: the mass of the lightest mesons is non--zero,
namely,
 \beq\label{Mgap}
 M_{\rm gap} \,=\,2\sqrt{2}\,\frac{u_m}{L^2}\,.\eeq
Note also that the meson masses are much smaller, by a factor $1/
\sqrt{\lambda}$, than the bare quark mass, \eqnum{mq}. This shows that in
this strong coupling limit the mesons are tightly bound: in the total
energy, the binding energy almost cancels the mass of the quarks.

%One can argue that such low--mass mesons are also rather small

\comment{ that the mesons should form multiplets transforming according
to the representations $(\ell/2,\ell/2)$ of SO(4)=SU(2)$\times$SU(2).
This degeneracy is indeed manifest\footnote{Actually, one finds an even
larger degeneracy: the modes furnish representations of the larger
symmetry group SO(5), but the respective degeneracy is not complete: not
all the modes in the same irreducible representation of SO(5) have the
same mass. See \cite{Kruczenski:2003be} for a discussion.} in the meson
spectrum computed in Ref.~\cite{Kruczenski:2003be}, which found the }

At finite temperature, the D7--brane feels the attraction exerted by the
black hole and thus is deflected towards the latter --- the stronger the
deviation, the shorter is the radial separation $u$ (or $\rho$) between
the two. This deflection becomes negligible towards the Minkowski
boundary ($r\to\infty$), where the profile function $R_v(r)$ approaches
the value $R_0$ that it would have (at any $r$) at $T=0$. More precisely,
for asymptotically large $r$ one finds
\cite{Babington:2003vm,Mateos:2006nu,mateos}
 \beq
 R_v(r)\,\simeq\,R_0 \,-\,\frac{c}{r^2}\,,\eeq
with $R_0$ related to the `bare' quark mass, as in \eqnum{mq}, and $c$ a
positive number proportional to the quark condensate.

On the other hand, closer to the black hole horizon ($\rho\sim 1$), one
finds two different types of behaviour
--- corresponding to two thermodynamically distinct phases separated by a
first--order phase transition ---, depending upon the ratio $M_{\rm
gap}/T=2\pi R_0$ between the (zero--$T$) mass gap and the temperature:
\texttt{(i)} for relatively large values of $R_0$, larger than a critical
value numerically found as $R_c\simeq 1.306$ \cite{mateos}, the
D7--branes close off above the black hole horizon (`low--temperature', or
`Minkowski embeddings'); \texttt{(ii)} for $R_0 < R_c$, the D7--branes
extend through the horizon (`high--temperature', or `black hole
embeddings')\footnote{There is a discontinuous jump in between these two
phases: with increasing temperature, the `tip' of the D7--brane which in
the Minkowski embeddings lies at $R_v(0)$ reaches an absolute minimum
value $R_v(0)\simeq 1.15$ corresponding to $R_0=R_c$ and then jumps into
the black hole horizon; see \cite{mateos} for details.}. In the gauge
theory, the most striking feature of this transition is the change in the
meson spectrum \cite{Mateos:2006nu} : in the low temperature phase, the
spectrum of mesons has a mass gap and the bound states are stable, so
like at $T=0$ \cite{mateos}; in the high temperature phase, there is no
mass gap and the mesonic excitations are unstable and characterized by a
discrete spectrum of quasinormal modes (\ie they have dispersion
relations with non--zero, and large, imaginary parts)
\cite{Hoyos:2006gb,starinets}.

For the reasons explained in the Introduction, in this paper we shall
restrict ourselves to the low--temperature phase, in which the mesons are
stable. The corresponding dispersion relations have been numerically
computed in Ref.~\cite{mateos}, at least within restricted regions of the
phase space. As expected, the spectrum shows deviations from both Lorentz
symmetry and supersymmetry, and these deviations become more and more
important with increasing temperature (for a given $M_{\rm gap}$). What
was perhaps less expected and, in any case, remarkable is the pattern of
the violation of the Lorentz symmetry by the spectrum: when increasing
the momentum $k$ of a given mode (\ie, for fixed values of $n$ and
$\ell$, which remain good `quantum numbers' also at finite temperature),
the `virtuality' $-Q^2\equiv \omega_{n\ell}^2(k) - k^2$ of that mode is
continuously decreasing, from time--like values ($-Q^2>0$) at relatively
low $k$ to space--like values ($-Q^2<0$) for sufficiently large $k$, in
such a way that, for asymptotically large $k$, the dispersion relation
approaches a {\em limiting velocity} which is strictly smaller than one:
 \beq\label{vm}
 \omega(k)\,\simeq\,v_0\,k\,\quad \mbox{with \ $v_0 \,< \,1$ \ as \
  $k\,\to\,\infty$}
 \,.\eeq
It has been furthermore noticed in the numerical analysis in
Ref.~\cite{mateos} that, with increasing $k$, the mode wavefunction
becomes more and more peaked near the bottom ($r\to 0$) of the D7--brane.
This led to the interesting suggestion, which was furthermore confirmed
by the numerical results, that the limiting velocity $v_0$ coincides with
the local velocity of light at $r\approx 0$, \ie $\rho\approx R_v(0)$ :
 \beq v_0\, \simeq\,  \left.
 \sqrt{-\frac{g_{tt}}{g_{zz}}}\right|_{r=0} =
 \frac{f(R_v(0))}{\tilde{f}(R_v(0))}\,.\labell{seer}\eeq
As we shall later argue in Sect.~\ref{sec:Mesons}, this identification
follows indeed from the respective EOM. Our analytic study will also
clarify other aspects of the dispersion relation, like the precise
conditions for the onset of the asymptotic behaviour \eqref{vm} and the
subleading corrections to it, which in particular contain the dependence
upon the quantum numbers $n$ and $\ell$. More generally, we shall be able
to construct piecewise analytic approximations for the dispersion
relation $\omega_{n\ell}(k)$ and also for the wavefunctions of the modes,
which will confirm the numerical findings in Ref.~\cite{mateos} and
provide further, analytic, insight into these results. Although, in our
analysis, we shall cover all kinematical domains in $k$ and thus provide
a global picture for the meson spectrum, our main focus will be on the
nearly light--light mesons with $\omega\simeq k$. Indeed, as we shall
explain in the next section, this regime is the only one to be relevant
for the deep inelastic scattering of the flavor current.

\section{Deep inelastic scattering off the ${\cal N}=2$ plasma}
\setcounter{equation}{0} \label{sec:DIS}

The ${\cal N}=2$ theory with $\nf$ flavors of equal mass has a global
$U(\nf) \simeq SU(\nf) \times U(1)_{\rm q}$ symmetry (describing flavor
rotations of the fields in the fundamental hypermultiplets), to which one
can associate $N_f^2$ conserved currents bilinear in the `quark'
operators (see Appendix A in \cite{starinets} for explicit expressions).
In particular, the current $J^\mu_{\rm q}$ corresponding to the diagonal
subgroup $U(1)_{\rm q}$ is associated with the conservation of the net
`quark' number (\ie the number of fundamental quarks and scalars minus
the number of antiquarks and hermitean conjugate scalars). By adding to
the theory a U(1)$_{\rm e.m.}$ gauge field $A_\mu$ minimally coupled to
this $J^\mu_{\rm q}$ current (with an `electromagnetic' coupling which is
arbitrarily small), one can construct a model for the electromagnetic
interactions and thus set up a {\em Gedanken} deep inelastic scattering
experiment which measures the distribution of the fundamental fields
inside the plasma. One can visualise this process as the exchange of a
virtual, space--like, `photon' (as described by the field $A_\mu$)
between the strongly coupled ${\cal N}=2$ plasma at finite temperature
and a hard lepton propagating through the plasma.

\subsection{Equations of motion in the D3/D7 brane model}
\label{sec:EOM}

Within the D3/D7 brane model, the flavor current $J^\mu_{\rm q}$ is dual
to an abelian gauge field $A_m$ living in the worldvolume of the
D7--brane, whose dynamics is encoded in the DBI action. According to the
gauge/gravity duality, the correlation functions of the operator
$J^\mu_{\rm q}$ are obtained from the `non--renormalizable' modes of the
field $A_m$, that is, the solutions to the classical EOM in the bulk of
the D7--brane which obey non--trivial (Dirichlet) boundary conditions at
the Minkowski boundary: as $r\to\infty$, the solution $A_m$ must approach
the U(1)$_{\rm e.m.}$ gauge field $A_\mu$ which acts as a source for the
current $J^\mu_{\rm q}$. This should be contrasted to the `normalizable'
modes dual to vector mesons, which must vanish sufficiently fast when
approaching the Minkowski boundary (see below for details).

In particular, the DIS cross--sections (or `structure functions') are
obtained from the (retarded) current--current correlator
 \beq
 \Pi^{\mu\nu}(q)\,\equiv\,i\int \rmd^4y\,\rme^{-iq\cdot y}\,\theta(y_0)\,
 \langle [J^\mu_{\rm q}(y), J^\nu_{\rm q}(0)]\rangle_T\,, \label{Pidef}  \eeq
where the brackets $\langle\cdots \rangle_T$ denote the thermal
expectation value in the  ${\cal N}=2$ plasma. To compute this two--point
function, it is enough to study the linearized EOM for the bulk field
$A_m$, \ie the same equations which determine the spectrum of the
low--lying vector mesons, but with different boundary conditions at
$r\to\infty$.

Specifically, the polarization tensor \eqref{Pidef} can be given the
following tensorial decomposition (in a generic frame) :
 \beq\label{PiTten}  \Pi_{\mu\nu}(q,T)=
  \left(\eta_{\mu\nu}-\frac{q_\mu q_\nu}{Q^2} \right)\Pi_1(x,Q^2)+
  \left(n_\mu-q_\mu \frac{n\cdot q}{Q^2}\right)
  \left(n_\nu-q_\nu \frac{n\cdot q}{Q^2}\right)\Pi_2(x,Q^2)\,,\eeq
where $\Pi_1$ and $\Pi_2$ are scalar functions, $n^\mu$ is the
four--velocity of the plasma in the considered frame, $Q^2=q_\mu q^\mu >
0$ is the (space--like) virtuality of the current, and
 \beq\label{xT}
       {x}\,\equiv\,\frac{Q^2}{-2(q\cdot n)T}
       %\,=\,\frac{Q^2}/{2\omega T}\,,
       \eeq
is the Bjorken variable for DIS off the plasma. Via the optical theorem,
the DIS structure functions are obtained as
 \beq  \label{F12T} F_1(x,Q^2)\,=\,\frac{1}{2\pi}\, {\rm Im} \,\Pi_1,\qquad
  F_2(x,Q^2)\,=\,\frac{-(n\cdot q)}{2\pi T}\, {\rm Im}\, \Pi_2\,. \eeq
In what follows, it will be convenient to compute the (boost--invariant)
structure functions by working in the plasma rest frame, where
$n^\mu=(1,0,0,0)$ and $q^\mu=(\omega,0,0,k)$, and therefore
$Q^2=k^2-\omega^2$ and $x={Q^2}/{2\omega T}$. However, one should keep in
mind that the physical interpretation of the results is most transparent
in the plasma `infinite momentum frame', \ie a frame in which the plasma
is boosted at a large Lorentz factor $\gamma\gg 1$. Then, the kinematic
invariants $Q^2$ and $x$ specify the transverse area ($\sim 1/Q^2$) and,
respectively, the longitudinal momentum fraction (equal to $x$) of the
plasma constituent (`parton') which has absorbed the space--like
`photon', and the structure functions represent parton distributions.

The piece of the DBI action which is quadratic in the gauge fields reads
(see \eg{} \cite{starinets})
 \beq S_8 = -\frac{(2 \pi \ell_s^2)^2}{4} T_{\rm D7} \nf \int \rmd^8 \sigma
 \sqrt{- g}\, g^{mp}g^{nq} F_{mn}F_{pq}\, , \label{SMax} \eeq
where $T_{\rm D7}=2\pi/(2\pi\ell_s)^8 g_s$, the space--time indices
$m,n,p,q$ run over the eight directions in the worldvolume of the
D7--brane, $g_{mn}$ is the induced metric on the D7--brane, and
$F_{mn}=\partial_m A_n-\partial_n A_m$. As already mentioned, the EOM
must be solved with the following boundary conditions
  \beq \label{pw}
   A_m(t,\bm{x},r\to\infty)\,\to\,A_\mu(t,\bm{x})
   \,=\,A_\mu^{(0)}\,\rme^{-i\omega t+ik z}\,, \eeq
which together with the fact that the equations are linear and
homogeneous in all the worldvolume directions but $r$ imply that the
solution $A_m$ is such that $A_r=A_{S^3}=0$ (\ie the radial and
$S^3$--components of the gauge field are identically zero) and the
remaining, four, components $A_\mu(t,\bm{x},r)$, with $\mu=t,x,y,z$, are
plane--wave in the Minkowski directions with $r$--dependent coefficients.
Since the gauge fields are independent of the coordinates on $S^3$, one
can reduce \eqnum{SMax} to an effective action in the relevant five
dimensions. The induced metric in these directions, that we denote as
$\tilde g_{mn}$, follows from Eqs.~(\ref{D3geom})--(\ref{change2}) as
 \beq
\rmd s^2(\tilde g) = \frac{1}{2} \left(\frac{u_0 \rho}{L}\right)^2
\left[-{f^2\over \tilde f}\,\rmd t^2 + \tilde{f} \rmd{\bm x}^2 \right] +
\frac{L^2}{\rho^2_v}\left(1 + \dot R_v^2\right)\rmd r^2\,, \label{D7ind}
 \eeq
where $\rho^2_v=r^2+R^2_v(r)$, $\dot R_v=\rmd R_v/\rmd r$, and we recall
that $R_v(r)$ is the profile of the D7--brane embedding. After
integrating over the coordinates on $S^3$, the action \eqref{SMax}
reduces to
 \beq S = -\frac{(2 \pi \ell_s^2)^2}{4} \Omega_3 T_{\rm D7} \nf \int \rmd t
\rmd^3x \rmd r \,\frac{\sqrt{- \tilde{g}}}{g_{\rm eff}^2(r)} \, \tilde
g^{mp} \tilde g^{nq} F_{mn}F_{pq} \,
 ,\label{SMax5} \eeq
where $\Omega_3 = 2\pi^2$, $m, n, \dots =t,x,y,z,r$, and
 \beq\label{geff}
\frac{1}{g_{\rm eff}^2(r)}\equiv \left(\frac{Lr}{\rho_v}\right)^3 \ \
\Longrightarrow \ \ \frac{\sqrt{- \tilde{g}}}{g_{\rm eff}^2(r)} \,=\,
 \frac{u_0^4}{4}\, r^2 f\tilde f\sqrt{1 + \dot R_v^2}\, .\eeq
Clearly, the EOM generated by the action \eqref{D7ind} read
 \beq\label{maxwell} \partial_m\left
 (\frac{\sqrt{- \tilde{g}}}{g_{\rm eff}^2(r)}
 \,\tilde g^{mp} \tilde g^{nq} F_{pq}\right)\,=\,0\,.
 \eeq

The propagation of the virtual photon along the $z$ axis introduces an
anisotropy axis in the problem, so the equations of motion look different
for the longitudinal ($\mu=t,z$) and respectively transverse ($\mu=x,y$)
components of the gauge field. In what follows we shall focus on the
transverse fields, $A_i$ with $i=x,y$, since from the experience with the
${\cal R}$--current \cite{HIM2,HIM3} we expect these fields to provide
the dominant contributions to the structure functions $F_{1,2}$ in the
high energy limit. Moreover, our final argument in Sect.~\ref{sec:resDIS}
will allow us to also reconstruct the flavor longitudinal structure
function $F_L=F_2-2xF_1$ from the corresponding one for the ${\cal
R}$--current. The relevant components of the field strength tensor are
$F_{ri}=\del_r A_i$, $F_{ti} =\del_t A_i\to -i\omega A_i$, and $F_{zi}
=\del_z A_i\to ik A_i$, and the equation satisfied by $A_i(r)$ reads
 \beq\label{EOMAi}
 \ddot A_i + \left[\del_r\ln\left(
 \frac{\sqrt{- \tilde{g}}}{g_{\rm eff}} \,\tilde g^{rr} \tilde g^{ii}
 \right)\right] \dot A_i \,+\,\frac{\tilde g^{zz}}{\tilde g^{rr}}
 \left(\frac{\tilde f^2}{f^2}\,\omega^2-k^2\right)A_i\,=\,0.
 \eeq
This equation must be solved with the Dirichelet boundary condition
\eqref{pw} at $r\to\infty$ together with the condition of regularity at
$r=0$. The solution to this boundary--value problem is a
`non--normalizable' mode, as opposed to the `normalizable' modes which
describe vector meson excitations of the D7--brane: the latter are the
solutions \eqnum{EOMAi} which vanish sufficiently fast (namely, like
$A_i(r)\sim 1/r^2$) when $r\to\infty$ \cite{Kruczenski:2003be,mateos}.

Once the `non--normalizable' solution is known as a (linear) function of
the boundary value $A_i^{(0)}$, the current--current correlator
\eqref{Pidef} is obtained, roughly speaking, by taking the second
derivative of the classical action (\ie the action \eqref{SMax5}
evaluated with that particular solution) with respect to $A_i^{(0)}$.
This procedure is unambiguous in so far as the euclidean (\ie
imaginary--time) correlators are concerned, but it misses the imaginary
part for the real--time correlators. Rather, the correct prescription for
computing the retarded polarization tensor \eqref{Pidef} reads (for the
case of $\Pi_{xx}=\Pi_{yy}=\Pi_1$) \cite{Son:2002sd,Herzog:2002pc}
 \beq\label{Pi1}
 \Pi_1(x,Q^2)\,=\,-\frac{\nf \nc T^2}{8}\,\left[r^3\,\frac
 {\del_r A_i(r,\omega,k)}{A_i(r,\omega,k)}\right]_{r\to\infty}\,,
 \eeq
where $i$ is either $x$ or $y$. The overall normalization factor reflects
the fact that the flavor current couples to $\nc\nf$ fundamental fields.
When using the above formula, the precise normalization at $r\to\infty$,
\ie the boundary value $A_i^{(0)}$, becomes irrelevant, as it cancels in
the ratio. Note that, in order to make use of \eqnum{Pi1}, it is enough
to know the solution in the vicinity of the the Minkowski boundary. But
to that aim, one generally needs to solve the EOM for arbitrary values of
$r$, since the second boundary condition is imposed at $r=0$.

In general, the coefficients in \eqnum{EOMAi} are rather complicated
functions, as visible on Eqs.~(\ref{D7ind}) and \eqref{geff}, and this
complication hinders the search for analytic solutions. However, how we
now explain, they can be considerably simplified without loosing any
salient feature by restricting ourselves to the very low temperature, or
very heavy meson, case $R_0\gg 1$, or $M_{\rm gap}\gg T$. This
restriction entails two important types of simplifications. The first one
refers to the `vacuum' profile $R_v(r)$, which in the general case is
known only numerically \cite{mateos}, but which becomes essentially flat
when $R_0\gg 1$. Indeed, in that case, the maximal deviation from the
asymptotic value $R_0$, namely (see App. A in Ref.~\cite{mateos}),
  \beq
  R_v(0)-R_0\,\simeq\,-\,\frac{1}{2R_0^7} \,\ll\,1\,,\eeq
is truly negligible, so one can use $R_v(r)\simeq R_0$ (and hence $\dot
R_v= 0$) at any $r$.

The second type of simplifications refer to the BH horizon at $\rho_0=1$
: when $R_0\gg 1$, the condition $\rho\gg 1$ is automatically satisfied
at any point within the worldvolume of the D7--brane. Then, the thermal
effects encoded in $f$ and $\tilde f$, which scale like $1/\rho^4$, cf.
\eqnum{ff}, can be safely neglected in all the terms in \eqnum{EOMAi}
{\em except} for the last one: indeed, within that term, the finite--$T$
deviations $1-f$ and $1-\tilde f$ are potentially amplified by the large
energy factor $\omega$. Specifically (with $\rho^2 =R_0^2+r^2$)
 \beq\label{ffomega}
 \frac{\tilde f^2}{f^2}\,\omega^2-k^2\,=\,(\omega^2-k^2)\,+\,
  \frac{4/\rho^4}{1-2/\rho^4}\,\omega^2\,\simeq\,
  -Q^2 \,+\,\frac{4}{\rho^4}\,
  \omega^2\,,\eeq
where we have also used $Q^2=k^2-\omega^2$.

To summarize, under the assumption that $R_0\gg 1$, the EOM for the
transverse gauge fields $A_i(r)$ takes a particularly simple form:
 \beq\label{EAi}
 \ddot A_i +
 \frac{3}{r}\,\dot A_i \,+
 \left(-\frac{2{\bar Q}^2}{\rho^4}
 \,+\,\frac{8\bar\omega^2}{\rho^8}\right)A_i\,=\,0\,,
 \eeq
where $\dot A_i=\rmd A_i/\rmd r$, $\rho^2 =R_0^2+r^2$, and we have
introduced the dimensionless variables
 \beq\label{barok}
 \bar\omega\,\equiv\,\omega\,\frac{L^2}{u_0}\,=\,
 \frac{\omega}{\pi T}\,,\qquad \bar k\,\equiv\,k\,\frac{L^2}{u_0}\,=\,
 \frac{k}{\pi T}\,,\qquad \bar Q^2\,\equiv\,\bar k^2-\bar\omega^2\,.
 \eeq
One should emphasize here that this condition $R_0\gg 1$ introduces no
loss of generality, neither for a study of the DIS process (in which case
we are anyway interested in $\omega,\,\bar Q\gg T$, and then the dominant
dynamics takes place at large radial distances $\rho\gg 1$
\cite{HIM2,HIM3}), nor for that of the meson spectrum (for which we shall
find results which are consistent with the numerical analysis in
Ref.~\cite{mateos}, although that analysis was performed for
$R_0\sim\order{1}$).

Although considerably simpler than the original equation \eqref{EOMAi},
the above equation is still too complicated to be solved exactly, except
in the special case $\bar Q=0$, to be discussed in
Sect.~\ref{sec:resDIS}. For more general situations, related to either
the meson spectrum or the problem of DIS, we shall later construct
analytic approximations. In preparation for that and in order to gain
more insight into the role of the various terms in \eqnum{EAi}, it is
useful to first consider a different but related problem, whose solution
is already known : this is the DIS of the ${\cal R}$--current
\cite{Polchinski:2002jw,HIM1,HIM2,HIM3}.

 \subsection{Some lessons from the ${\cal R}$--current}
 \label{sec:R}

The ${\cal R}$--current is a conserved current associated with one of the
U(1) subgroups of a global SU(4) symmetry of the ${\cal N}=4$ SYM theory.
The respective operator is bilinear in the massless, adjoint, fields of
${\cal N}=4$, and remains conserved even in the presence of the
fundamental hypermultiplets (\ie in ${\cal N}=2$ theory), because of the
probe limit $g^2\nf\ll 1$. The supergravity field dual to the ${\cal
R}$--current is, once again, a gauge field $A_\mu$, whose dynamics
however is not anymore restricted to the worldvolume of the D7--brane ---
rather, this field can propagate everywhere in the AdS$_5\times S^5$
Schwarzschild space--time, in particular, it can fall into the black
hole. Because of that, the D7--brane plays no role in the case of the
${\cal R}$--current, so the following discussion applies to both ${\cal
N}=4$ and ${\cal N}=2$ theories (with $\nf\ll\nc$, of course).

For a space--like ${\cal R}$--current with high virtuality $Q\gg T$ and
for large radial coordinates $\rho\gg \rho_0$, the dynamics of the dual
${\cal R}$--field $A_i(r)$ is described by an equation similar to
\eqnum{EAi}, but where the variables $\rho$ and $r$ are now identified
with each other (since $R_0$ plays no role in this case). That is,
 \beq\label{EAiR}
 \ddot A_i +
 \frac{3}{\rho}\,\dot A_i \,+
 \left(-\frac{2{\bar Q}^2}{\rho^4}
 \,+\,\frac{8\bar\omega^2}{\rho^8}\right)A_i\,=\,0\qquad
 \mbox{(${\cal R}$--current)}\,,
 \eeq
where now $\dot A_i=\rmd A_i/\rmd \rho$ and it is understood that
$\rho\gg 1$.

The dynamics is driven by the competition between the two terms inside
the brackets in \eqnum{EAiR}. The first term, proportional to $Q^2$, acts
as a potential barrier which opposes to the progression of the field
towards the interior of AdS$_5$ : by itself, this would confine the field
near the Minkowski boundary, at large radial distances $\rho\gtrsim
\rho_Q\equiv \bar Q$. The second term, proportional to $\omega^2$, is
present only at finite temperature (as manifest from its derivation in
\eqnum{ffomega}) and it represents the gravitational attraction between
the gauge field and the BH. For sufficiently small values of $\rho$,
smaller than $\rho_c\equiv ({2\bar\omega/\bar Q})^{1/2}$, this attraction
overcomes the repulsive barrier $\propto Q^2$, and then the overall
potential becomes attractive. However, unless the energy $\bar\omega$ is
high enough, this change in the potential has no dynamical
consequences\footnote{We here ignore the tunneling phenomenon, which is
exponentially suppressed \cite{HIM2,Mueller:2008bt}.}, because the field
is anyway stuck near the Minkowski boundary and thus cannot feel the
attraction. Clearly a change in the dynamics will occur when the energy
is so high that $\rho_c\gtrsim \rho_Q$, which requires $\bar\omega
\gtrsim \bar Q^3$, or, in physical units, $\omega \gtrsim Q^3/T^2$. When
this happens, the potential barrier at $\rho\to\infty$ cannot prevent the
gauge field to penetrate (through diffusion; see the discussion in
Sect.~\ref{sec:TIME}) down to the attractive part of the potential at
$\rho\lesssim\rho_c$, and from that point on, the potential barrier plays
no role anymore. Hence, the dynamics at radial distances $1\ll
\rho\lesssim \rho_c$ is controlled by the even simpler equation
 \beq\label{HighR}
 \ddot A_i +
 \frac{3}{\rho}\,\dot A_i \,+
 \,\frac{8\bar\omega^2}{\rho^8}\,A_i\,=\,0\qquad
 \mbox{(${\cal R}$--current \& high energy)}\,,
 \eeq
which does not involve $\bar Q$ and thus is formally the same as the
equation describing a light--light current. This equation can be exactly
solved in terms of Airy functions. The general solution reads (up to an
overall normalization, which is irrelevant)
 \beq\label{AiAiry}
 A_i(\rho)\,=\,c{\rm Ai}(-\xi)\,+\,{\rm Bi}(-\xi)\,,
 \qquad\xi\,\equiv\,\left(\frac{\sqrt{2}\bar\omega}{\rho^3}
 \right)^{2/3}\,,\eeq
valid for $(\bar Q/ \bar\omega^{1/3})\lesssim\xi\ll \bar\omega^{2/3}$.
This involves one unknown coefficient $c$ which can be fixed, in
principle, by matching onto the corresponding solution at smaller
distances $\rho\sim \order{1}$, which in particular obeys the appropriate
boundary condition at $\rho=\rho_0=1$. This boundary condition is rather
clear on physical grounds: the gauge field can be only absorbed by the
BH, but not also reflected, hence the near--horizon solution must be a
{\em infalling} wave \cite{Son:2002sd,Son:2007vk}, \ie a field which with
increasing time approaches the horizon.

The solution near $\rho=1$ obeying this boundary condition can be
explicitly computed, and its matching onto \eqnum{AiAiry} can indeed be
done \cite{HIM2}, but it turns out that this actually not needed for the
purpose of computing the DIS structure function: the problem of the
${\cal R}$--current offers an important simplification, which is worth
emphasizing here, since the same simplification appears for the flavor
current in the high--temperature case (the `black hole embedding')
\cite{Bayona:2009qe}, but not also in the low--temperature, or
`Minkowski', embedding of interest for us here. Namely, the infalling
boundary condition can be enforced not only near the BH horizon, but also
at much larger values of $\rho$, where \eqnum{AiAiry} applies. This is so
since there is no qualitative change in the shape of the potential at any
intermediate point in the range $1< \rho\ll\rho_c$ which could give rise
to a reflected wave.

The last observation allows us to identify $c=i$ in \eqnum{AiAiry}:
indeed, consider this approximate solution for $\rho \ll
\bar\omega^{1/3}$, or $\xi\gg 1$, where one can resort on the asymptotic
expansions for the Airy functions. Using \eqnum{AiAiry} with $c=i$, one
obtains
 \beq\label{AiRsol}
 A_i(t,z,\rho)\,\simeq\,\frac{1}{\sqrt{\pi}\xi^{1/4}}
 \,\exp\left
 \{-i\omega t+ik z + i\frac{2}{3}\,\xi^{3/2}+ i\frac{\pi}{4}
 \right\}\quad\mbox{for}\quad 1\,\ll\,\rho\,\ll\,\bar\omega^{1/3}
 \,,\eeq
which is indeed an infalling wave.

Now that the coefficient $c$ has been fixed, one can use the approximate
solution \eqref{AiAiry} for relatively small values of $\xi$ and compute
the current--current correlator according to \eqnum{Pi1}. (One can adapt
\eqnum{Pi1} to the ${\cal R}$--current by multiplying its r.h.s by a
factor $\nc/4\nf$.) Specifically, \eqnum{AiAiry} is still correct for
$\xi\sim \bar Q/ \bar\omega^{1/3}\ll 1$, where one can use the
small--$\xi$ expansions for the Airy functions, and thus deduce
\cite{HIM2}
 \beq\label{F1R}
  F_1\,=\,\frac{3\nc^2T^2}{16\Gamma^2(1/3)}\left(\frac{\bar\omega}{6}
   \right)^{{2}/{3}}\,,\qquad F_2\,\sim\,{x}F_1\,
   \sim\,x\nc^2Q^2\left(\frac{T}{xQ}
   \right)^{{2}/{3}}
   \,,\eeq
where in the last estimate we indicated the parametric dependencies of
the structure functions upon the variables relevant for DIS\footnote{One
can show that $F_L\equiv F_2-2xF_1$ is parametrically of the same order
as $xF_1$ when $x\sim x_s$, but it is relatively negligible when $x\ll
x_s$ \cite{HIM2}.}. As it should be clear from the previous discussion,
these results hold for sufficiently high energy, $\omega \gtrsim
{Q^3}/{T^2}$, a condition which can be rewritten in terms of the DIS
variables $x$ and $Q^2$ as
 \beq\label{Qs}
  x\,\equiv
 \,\frac{Q^2}{2\omega T}\ \lesssim\ x_s(Q)\,\equiv\,
 \frac{T}{Q}\,,\qquad\mbox{or}\qquad
 Q\ \lesssim\ Q_s(x)\,\equiv \,\frac{T}{x}\,.\eeq
On the other hand, for larger values of Bjorken--$x$, $x \gg T/Q$, or
higher virtualities $Q\gg Q_s$, the structure functions are exponentially
small (since generated through tunelling). This strong suppression of the
structure functions at large values of $x$ and/or $Q^2$ implies the
absence of point--like constituents in the strongly coupled plasma
\cite{Polchinski:2002jw,HIM1,HIM2,HIM3}. The critical value $Q_s(x)\sim
T/x$ is known as the {\em saturation momentum}, since \eqnum{F1R} is
consistent with a parton picture in which partons occupy the phase space
at $Q\lesssim Q_s(x)$ with occupation numbers of $\order{1}$ \cite{HIM2}.

Returning to the flavor current of interest here, let us now identify the
similarities and the differences with respect to the problem of the
${\cal R}$--current, that we have just discussed.

In the {\em high--temperature} case, where the tip of the D7--brane
enters the BH horizon, there are no serious conceptual differences with
respect to the ${\cal R}$--current. For $r\gg 1$, \eqnum{EAi} is still
valid, so the large--$r$ dynamics is exactly the same as discussed in
relation with \eqnum{AiAiry}. At smaller $r\sim\order{1}$, the EOM
becomes more complicated (in particular because of the $r$--dependence of
the profile function $R_v(r)$, which is non--trivial in that
high--temperature case), but there is no ingredient in the dynamics which
could prevent the fall of the flavor field $A_i$ into the BH. Hence, the
appropriate boundary condition at $\rho=1$ (the tip of the D7--brane) is
still the infalling one, and moreover this condition can again be
enforced ar large $r\gg 1$, where \eqnum{AiAiry} applies. As before, this
condition fixes $c=i$, thus finally yielding the same result for the DIS
structure function as in \eqnum{F1R}, except for the overall
normalization:
 \beq\label{F1flHT}
  F_1\,\simeq\,%\frac{F_2}{2x}\,=\,
  \frac{3\nc\nf T^2}{4\Gamma^2(1/3)}\left(\frac{\bar\omega}{6}
   \right)^{{2}/{3}}\qquad\mbox{(flavor current
   in the BH embedding)}
   \,.\eeq
This is indeed the result found in \cite{Bayona:2009qe}. In particular,
the saturation momentum for the flavor current (in this high--temperature
regime, at least) is exactly the same as for the ${\cal R}$--current, cf.
\eqnum{Qs}, since fully determined by the current interactions with the
BH.

Consider now the {\em low--temperature} phase, which is the most
interesting case for us here. For the DIS problem, it is natural to
assume that $Q\gtrsim M_{\rm gap}\gg T$, or $\bar Q\gtrsim R_0\gg 1$. The
situation near the Minkowski boundary will be quite similar to that for
the ${\cal R}$--current: At relatively low energies $\bar\omega \ll \bar
Q^3$, there is a potential barrier at $\rho\gtrsim \bar Q
> R_0$, which however disappears at larger energies $
\bar\omega \gtrsim \bar Q^3$. When this happens, the flavor field can
penetrate all the way within the worldvolume of the D7--brane. However,
this worldvolume ends up at $\rho=R_0\gg \rho_0$, so there is clearly no
possibility for this field to fall into the BH. Accordingly, the
infalling boundary conditions do not apply here, but rather must be
replaced by the condition of regularity at $\rho=R_0$ (the tip of the
D7--brane). In order to establish the fate of the high--energy current,
we therefore cannot make the economy of actually solving \eqnum{EAi} for
all the values of $\rho$ down to $R_0$. Still, there is an important
`technical' simplification which occurs at high energy: then, the
potential barrier $\propto \bar Q^2$ plays no role anymore, so it is
sufficient to consider the $\bar Q=0$ version of \eqnum{EAi}. This
equation also determines the spectrum of the light--like mesons in the
plasma and, as we shall see, these two problems
--- the flavor DIS and the light--like mesons --- are indeed strongly
related to each other.

\section{Meson spectrum at low temperature}
%: from time--like to space--like}
\setcounter{equation}{0} \label{sec:Mesons}

In this section we shall construct piecewise approximations to the
spectrum of the meson excitations in the low temperature phase, or
Minkowski embedding, with the purpose of clarifying some global
properties of the spectrum numerically obtained in Ref.~\cite{mateos} and
exhibited in Fig.~\ref{fig:dispersion}. In particular, we shall follow
the transition of the dispersion relation of a given mode from time--like
to space--like with increasing momentum $k$, and thus identify the
`critical' momentum $k_n$ at which the mode with radial quantum number
$n$ crosses the light--cone. Also we shall recover previous analytic
results in the literature which concentrated on special limits, like
zero--temperature \cite{Kruczenski:2003be} or very high momentum
\cite{Ejaz:2007hg}. The particular case of a light--like mode
($\omega_n(k)=k$) will be given further attention in the next section,
where we shall construct the exact respective solutions for both
normalizable and non--normalizable modes, with the purposes of
understanding DIS.

\subsection{Equation of motion in Schr\"odinger form}

For the subsequent analysis, it is convenient to change the definition of
the radial coordinate once again, in such a way that the infinite
interval $0\le r < \infty$ be mapped into the compact interval $0\le
\zeta \le 1$. Here $\zeta$ is defined as
 \beq\label{zeta}
 \zeta\,\equiv\,\frac{r^2}{\rho^2(r)}\,=\,\frac{r^2}{R_0^2+r^2}\,,\eeq
so in particular the Minkowski boundary corresponds to $\zeta=1$. Then
\eqnum{EAi} becomes
 \beq\label{EPhi}
 \left(\del_\z^2 \,+\,\frac{2}{\z}\del_\z \,-\,
 \frac{\kappa^2}{4\z(1-\z)}\,+\,\frac{4\Omega^2(1-\z)}{\z}\right)
 \Phi(\z)\,=\,0\,,
 \eeq
where we have replaced the name of the function by $\Phi$, for more
generality: indeed, Eqs.~\eqref{EAi} or \eqref{EPhi} apply not only to
transverse vector mesons, but also to the pseudoscalar mesons
corresponding to small fluctuations in the azimuthal angle $\phi$ (the
angle in the 89--plane transverse to the D7--brane; recall that the
`vacuum' embedding corresponds to $\phi=0$). Furthermore, we have defined
 \beq\label{kapomeg}
 \kappa^2\,\equiv\,\frac{2\bar Q^2}{R_0^2}\,,\qquad \Omega^2
 \,\equiv\,\frac{\bar\omega^2}{2R_0^6}\,,
 \eeq
where the virtuality $\bar Q^2=\bar k^2-\bar\omega^2$ can now take any
sign (and thus the same is true for $\kappa^2$). It is furthermore
convenient to rewrite \eqnum{EPhi} in the form of a Schr\"odinger
equation, \ie to remove the term involving the first derivative; this can
be done by writing
 \beq\label{Psi}
 \Phi(\z)\,=\,\frac{1}{\z}\,\psi(\z)\,.\eeq
The corresponding ``Schr\"odinger equation'' reads
 \beq\label{Epsi}
 \left(-\del_\z^2 \,+\,V_{\ell}(\z)\right)
 \psi(\z)&\,=\,&0\,,\\*[0.2cm]
  V_{\ell}(\z)\,\equiv\,\frac{\ell(\ell+2)}
 {[2\z(1-\z)]^2} &\,+\,&
 \frac{\kappa^2}{4\z(1-\z)}\,-\,\frac{4\Omega^2(1-\z)}{\z}\,,
 \label{Vl}\eeq
where we have allowed for one further generalization by adding to the
potential the term corresponding to a generic value $\ell$, with
$\ell=0,1,2,\dots$, for the angular `quantum number' corresponding to
rotations around the $S^3$--sphere internal to the D7--brane. (Such
rotations cannot be excited by either the flavor or the ${\cal
R}$--current considered in the previous section, so $\ell=0$ in the case
of DIS. But modes with non--zero $\ell$ can be excited by other operators
in the boundary gauge theory, which are charged under the global SO(4)
symmetry of the fundamental hypermultiplets
\cite{Kruczenski:2003be,mateos,starinets}.) Since the radial coordinate
$\z$ terminates at $\z=1$, it is understood that the potential becomes an
infinite wall at that point; given the structure of \eqnum{Vl}, this
additional constraint has no consequence except in the limiting case
where $\ell=0$ and $\kappa^2=0$. The zero temperature case is obtained by
formally taking $\Omega=0$ in \eqnum{Vl}.

Since we are interested in the normalizable modes describing mesons, we
shall look for solutions $\psi_{\ell}(\z)$ to \eqnum{Epsi} obeying the
following boundary conditions \cite{Kruczenski:2003be} :
 \begin{equation}\label{philimits}
    \psi_{\ell}(\z)\,\propto\,
    \begin{cases}
        \displaystyle{\z^{\frac{\ell+2}{2}}} &
        \text{ for\,  $\z\,\to\, 0$}
        \\*[0.2cm]
        \displaystyle{
        (1-\z)^{\frac{\ell+2}{2}}} &
        \text{ for\,  $\z \,\to\,1$}
        .
    \end{cases}
 \end{equation}
In what follows we would like to follow the change in the dispersion
relation when increasing the meson momentum $k$ for fixed quantum numbers
(\ie for a given mode). This study will drive us through different
regimes in terms of the variables $\kappa^2$ and $\Omega^2$. The
potential $V(\z)\equiv V_{\ell=0}(\z)$ in these various regimes is
illustrated in Figs.~\ref{fig:Vab} and \ref{fig:Vcd}.

\FIGURE[t]{
\centerline{%\hspace*{-3.cm}
\includegraphics[width=0.5\textwidth]{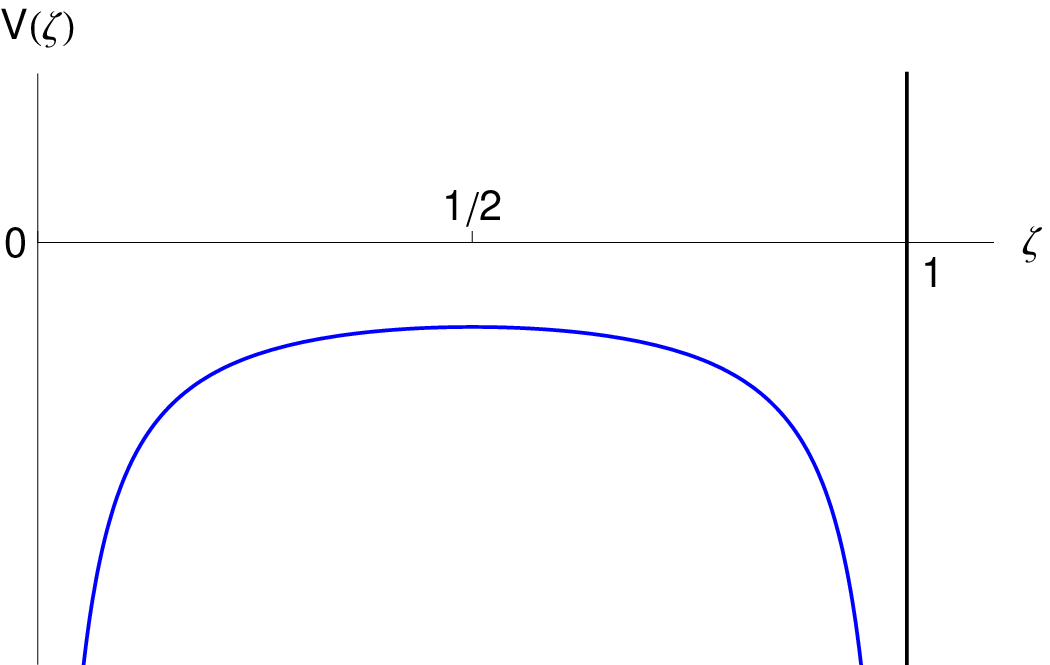}
\includegraphics[width=0.5\textwidth]{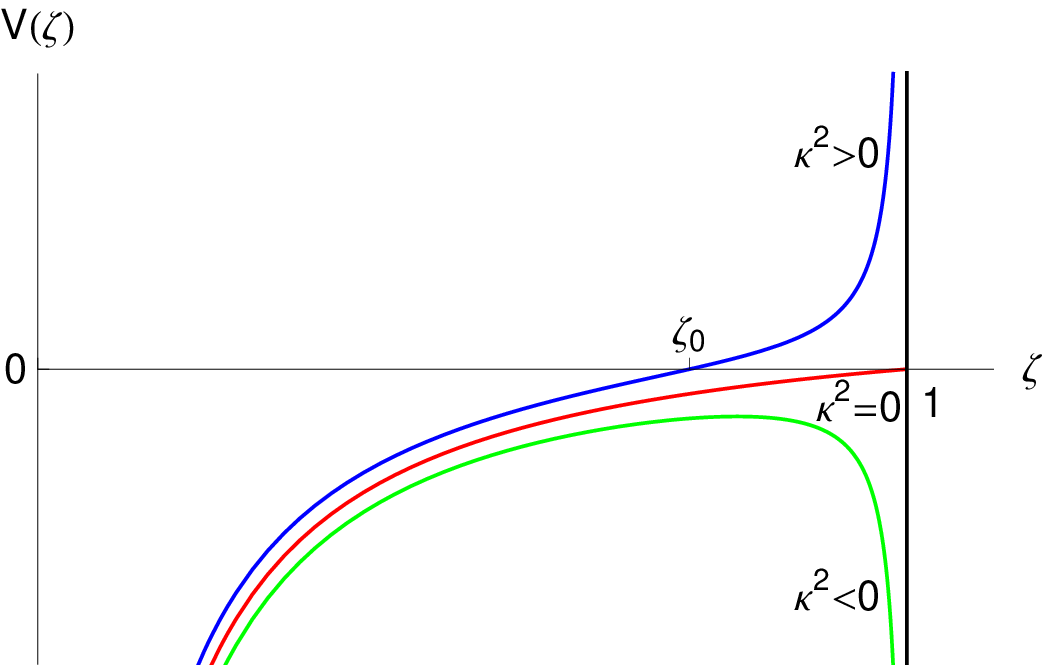}
} \caption{\sl The potential $V(\z)$ in \eqnum{Vl} for $\ell=0$.
%\eqnum{Vl} for $\ell=0$.
Left: the case where $\Omega\ll \kappa$ (as relevant at zero temperature,
or for finite temperature but relatively small momenta $\bar k\ll
nR_0^3$). Right: the case where $|\kappa|\ll \Omega$ (this corresponds to
modes with momenta $\bar k\sim nR_0^3$). \label{fig:Vab}}}

\FIGURE[t]{
\centerline{%\hspace*{-3.cm}
\includegraphics[width=0.5\textwidth]{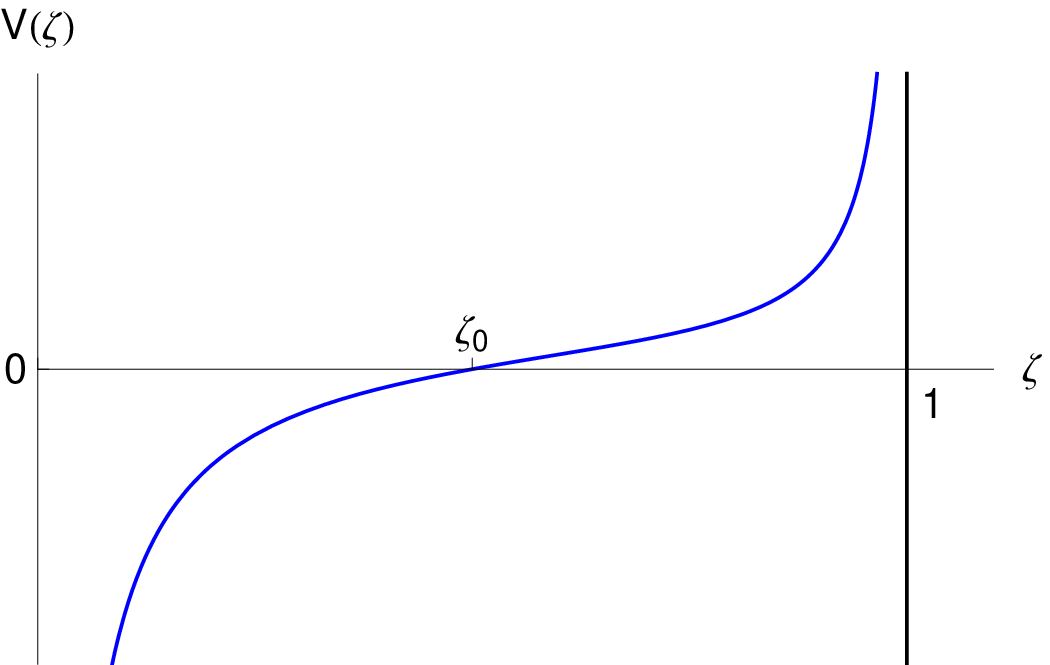}
\includegraphics[width=0.5\textwidth]{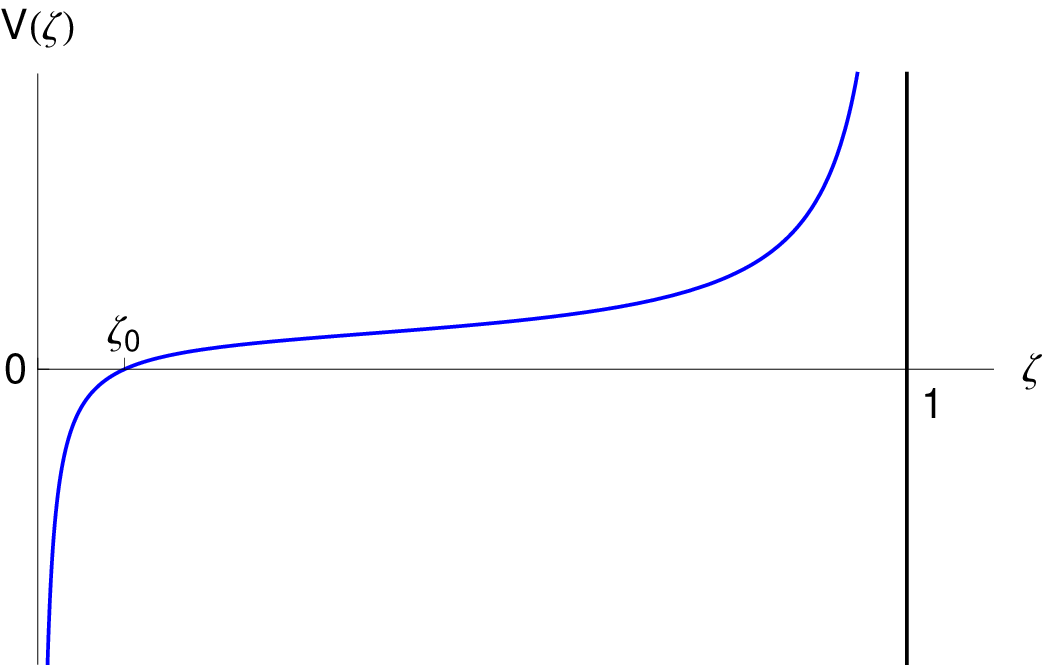}
 } \caption{\sl  The potential $V(\z)$ in \eqnum{Vl} for $\ell=0$
and relatively large, space--like, virtualities, with $\kappa < 4\Omega$.
Left: $\kappa = 2\Omega$. Right: $\kappa\simeq 4\Omega$ (the limiting
velocity regime). \label{fig:Vcd}}}

\subsection{The low momentum regime: time--like dispersion relation}

%\bigskip
%\texttt{(i) The low momentum regime: time-like dispersion relation}
%\bigskip

When the momentum $k$ is sufficiently small (see \eqnum{TLcond} below for
the precise condition), the dispersion relation is time--like
($\kappa^2<0$) and it is such that the last term, proportional to
$\Omega^2$, in the potential becomes negligible (so that the potential
has the symmetric shape shown in Fig.~\ref{fig:Vab} left). Then
\eqnum{EPhi} is formally the same as at zero temperature and the
corresponding solutions are known exactly \cite{Kruczenski:2003be}.
Namely, the solution obeying the right boundary condition at $\z=0$, cf.
\eqnum{philimits}, reads (the overall normalization is chosen for
convenience)
 \beq\label{psismallk}
 \psi_{\ell}(\z)&\,=\,&[\z(1-\z)]^{\frac{\ell+2}{2}}
 \,F(a,b;\ell+2;\z)\,,\\*[0.2cm] \nonumber
 \mbox{with}\quad
 a&\,\equiv\,&\frac{1}{2}\big(2\ell + 3+\sqrt{1+\mu^2}\big)\,,\quad
 b\,\equiv\,\frac{1}{2}\big(2\ell + 3-\sqrt{1+\mu^2}\big)\,,
 \eeq
where $F(a,b;c;x)$ is the usual hypergeometric function, also denoted as
${}_2F_{1}(a,b;c;x)$, which obeys $F (a,b;c;0)=1$ (see \eg{} Chapter 15
in \cite{AbramStegun}) and we have set $-\kappa^2\equiv \mu^2>0$. For
\eqnum{psismallk} to also obey the correct boundary condition at $\z=1$,
the hypergeometric function must be regular at that point, which for the
indicated values of the parameters $a,b$ and $c=\ell+2$ requires the
hypergeometric series to terminate \cite{AbramStegun}. That is, $b=-n$
with $n=0,1,2,...$, and then $F(a,-n;c;x)$ is a polynomial of degree $n$
in $\z$. This condition yields the vacuum--like (\ie $T=0$) spectrum in
\eqnum{T0spectrum}, that is,
 \beq\label{barM}
 \mu^2(n,\ell)\,\equiv\, \frac{2(\bar\omega_{n\ell}^2(k) - \bar k^2)}
 {R_0^2}\,=\,4(n+\ell+1)(n+\ell+2)\,.\eeq
At finite temperature, \eqnum{barM} remains a good approximation so long
as one can neglect the term $\propto\Omega^2$ in the potential, that is,
for $\Omega^2_n\ll \mu^2_n$ or, equivalently (recall \eqref{kapomeg}),
$\bar\omega\ll nR_0^3$. Since we also assume $R_0\gg 1$, it is clear that
this condition is satisfied up to relatively high values of the momentum
$k$, namely so long as
 \beq\label{TLcond}
 \bar k\,\ll\,nR_0^3\,,\qquad\mbox{or}\qquad k\,\ll\,n\,(M_{\rm
 gap}^3/T^2)\,.\eeq
Note that we include the radial quantum number within parametric
estimates, \eg, $\mu_n\sim n$ or $\bar\omega_n\sim nR_0$, since we shall
be also interested in large values $n\gg 1$ (whereas $\ell$ will never be
too large). Although, for definiteness, we refer to the kinematical
domain \eqref{TLcond} as the `low--momentum regime', it is clear that,
towards the upper end of this domain, the momenta are so large that $k\gg
M_n \sim n M_{\rm gap}$ and thus the dispersion relation becomes nearly
light--like.

Still within this low--momentum regime, it is easy to see that the main
effect of the ``finite--$T$'' term $\propto\Omega^2$ in the potential
\eqref{Vl} is to decrease the meson virtuality as compared to its
``zero--$T$'' value \eqref{barM}. Indeed, a simple estimate for this
effect is obtained by replacing $\mu^2 \to \mu^2 +16 \Omega^2$ in the
l.h.s. of \eqnum{barM} (this procedure overestimates the correction when
$\z\simeq 1$, but it should be correct at least qualitatively); hence the
corrected virtuality reads
 \beq\label{barMc}
 \mu^2(k,n,\ell)\,\approx\,\mu^2_{(0)}(n,\ell)\,-\,16\Omega^2_{n\ell}\,,
 \quad\mbox{with}\quad \mu^2_{(0)}(n,\ell)\,\equiv\,
 4(n+\ell+1)(n+\ell+2)\,.\eeq
In particular, for non--relativistic momenta ($k\ll M_n\sim n M_{\rm
gap}$), the above dispersion relation can be expanded out as (the quantum
numbers are kept implicit)
 \beq\label{nonrel}
 \omega(k)\,\approx\,M_{\rm rest}\,+\,\frac{k^2}{2M_{\rm kin}}\,,\quad
 \mbox{with}\quad M_{\rm rest}\approx M_{(0)}(1-2/R^4_0)\,,\quad
 M_{\rm kin}\approx \frac{M_{(0)}}{1-2/R^4_0}\ ,\eeq
and therefore $M_{\rm rest}M_{\rm kin}\approx M_{(0)}^2$. ($M_{(0)}$
denotes the ``zero--$T$'' meson mass, as given by \eqnum{T0spectrum} or
\eqref{barM}.) The estimates \eqref{nonrel} are in fact in agreement with
the respective numerical findings in Ref.~\cite{mateos} (see the
discussion of Eq.~(4.45) there).

\subsection{The intermediate momentum regime: light--like dispersion
relation} \label{sec:LL}

%\bigskip
%\texttt{(ii) The intermediate momentum regime: light-like dispersion
%relation}
%\bigskip

With further increasing $k$, the energy $\omega_{n}$ of the mode $n$ is
also increasing and the last term, proportional to $\Omega^2$, in the
potential \eqref{Vl} becomes more and more important. Since, at the same
time, the virtuality $\mu_n^2$ of the mode is decreasing, it should be
clear that for sufficiently large $k$ --- namely, when $\bar k\sim
nR_0^3$ --- one enters a regime where $\Omega^2_n\gg \mu_n^2$ and then
the roles of the respective terms in the potential are interchanged: the
term in $\Omega^2$ becomes the dominant one, while that in $\mu^2$
represents only a small correction. Then the mode $n$ is nearly
light--like and in fact it crosses the light cone (\ie its virtuality
$\mu_n^2(k)$ changes sign) when varying $k$ within this domain. We have
not been able to analytically follow this transition, but the fact that
it actually happens is quite obvious by inspection of the shape in the
potential in this regime. This is shown in Fig.~\ref{fig:Vab} right for
the three cases of interest: (a) $\kappa^2\equiv -\mu^2$ is negative but
small, (b) $\kappa^2=0$, and (c) $\kappa^2$ is positive but small.

Namely, consider the genuine Schr\"odinger equation associated to this
potential, that is,
  \beq\label{Schrod}
 \left(-\del_\z^2 \,+\,V(\z)\right)
 \psi(\z)\,=\,E\psi(\z)\,,\eeq
where $E$ is the energy of a bound state. Given the shape of the
potential in Fig.~\ref{fig:Vab} right, it is clear that bound states with
both positive and negative energies will exist for all the three cases
aforementioned. It is furthermore clear that, with increasing $\Omega^2$
at fixed $\kappa^2$ the potential becomes more and more attractive, so
some of the bound states will cross from positive to negative energies.
This means that, for any fixed value of $\kappa^2$, there exist
corresponding values of $\Omega^2$ such that the respective bound states
have $E=0$. These are, of course, the meson modes that we are interested
in.

This Schr\"odinger argument also suggest the use of the semi--classical
WKB method for computing the meson spectrum. Given the shape of the
potential this should be a reasonable approximation at least for
sufficiently large numbers $n\gg 1$ (we set $\ell=0$ for simplicity). The
Bohr--Sommerfeld quantization condition for the mode $n$ with energy
$E_n=0$ reads
 \beq\label{BS}
 n\pi\,=\,\int_0^{\z_0}\rmd\z\,\sqrt{-V(\z)}\,,\eeq
where $\z_0$ is the turning point in the potential, that is, $\z_0=1$
when $\kappa^2\le 0$ and $\z_0=1-\kappa/4\Omega$ when $\kappa^2>0$. When
$\kappa=0$, the integral is straightforward and yields
 \beq\label{WKBn}
 \Omega_n\,\approx\,n\,,\qquad\mbox{or}\qquad
 \bar\omega_n\,\approx\,\sqrt{2}\,n\,R_0^3 \,.\eeq
That is, the mode $n$ crosses the light--cone at $k=k_n$ with $\bar
k_n\approx \sqrt{2}n R_0^3$. As we shall see in Sect.~\ref{sec:resDIS},
this is indeed the correct result when $n\gg 1$.

An interesting property of the spectrum near the light--cone, which will
play an important role in our subsequent study of DIS (see
Sect.~\ref{sec:resDIS}) and can be also understood on the basis of
\eqnum{BS}, is the extreme sensitivity of the dispersion relation to
changes in the virtuality $\kappa^2$ around $\kappa=0$ : so long as
$|\kappa^2|\ll \Omega^2$, a small change in $\kappa^2$ entails a large
change in $\Omega^2$. Before we explain the origin of this property, let
us first use \eqnum{BS} to render it more specific. Consider the
space--like case $\kappa^2>0$ for definiteness, and denote
$\varepsilon\equiv \kappa/4\Omega\ll 1$, so that the turning point lies
at $\z_0=1-\varepsilon$. Changing the integration variable according to
$x\equiv 1-\z$, we can successively write
 \beq\label{BSLL}
 \frac{n\pi}{2\Omega}&\,=\,&\int_\varepsilon^1\rmd x\sqrt{\frac
 {x^2-\varepsilon^2}{x(1-x)}}
 \,\simeq\,\frac{\pi}{2} +\int_0^1\frac{\rmd x}{ \sqrt{x(1-x)}}
 \left(\Theta(x-\varepsilon)
 \sqrt{x^2-\varepsilon^2} - x\right)\\*[0.2cm] \nonumber
 &\,\simeq\,&\frac{\pi}{2}\, -\, \varepsilon^{3/2}
 \int_0^\infty\frac{\rmd \lambda}{\sqrt{\lambda}}
 \left(\lambda-\Theta(\lambda-1)\sqrt{\lambda^2-1}\right),
 % \,=\,\frac{\pi}{2}\, -\, C\varepsilon^{3/2}
 \eeq
where we have observed that, after subtracting the dominant contribution
$\pi/2$ to the first integral, the subtracted integral is dominated by
its lower limit $x=\varepsilon$; this allowed us to perform the
simplifications in the second line, where we denoted
$x\equiv\lambda\varepsilon$. The final integral multiplying
$\varepsilon^{3/2}$ is clearly a positive number of $\order{1}$. One can
combine together the space--like and time--like cases into the following
formula
 \beq\label{smallk}
  n \approx \Omega_n \,-\,{\rm sgn}(\kappa^2)C
  \sqrt{\frac{|\kappa|^3}{\Omega_n}}
 \qquad\mbox{for}\qquad|\kappa|\,\ll\, \Omega\,,\eeq
where $C$ is a positive constant and ${\rm sgn}(x)=\Theta(x)-\Theta(-x)$
is the sign function. This formula shows that, when moving away from the
light--cone, say, towards space--like virtualities, the energy of the
mode grows by a substantial amount $\Delta \Omega_n\sim 1$ for a
relatively modest increase in the virtuality, from $\kappa=0$ to
$\kappa\sim \Omega_n^{1/3}\ll \Omega_n$. In physical units, we change
$\bar\omega_n$ by a large amount $\Delta\bar\omega_n\sim R_0^3 \gg 1$
when increasing $\bar Q$ from zero to $\bar Q\sim n^{1/3}R_0\sim
\bar\omega_n^{1/3}$.

This strong sensitivity of the dispersion relation to $\kappa^2$ around
$\kappa=0$ can be traced back to the behavior of the potential near the
turning point $\z_0$ (recall that it is this turning point which controls
the $\kappa$--dependence of the integral in \eqnum{BS}). Namely, from
\eqnum{Vl} we deduce (for positive $\kappa$ with $\kappa\ll\Omega$)
 \beq \frac{\del V}{\del\kappa^2}\Bigg |_{\z=\z_0}\,\simeq\,
 \frac{\Omega}{\kappa}\,\gg\,1\,,\quad\mbox{and}\quad
  \frac{\del (- V)}{\del\Omega^2}\Bigg |_{\z=\z_0}\,\simeq\,
  \frac{\kappa}{\Omega}\,\ll\,1\,.\eeq

\subsection{The high momentum regime: space--like dispersion relation}

%\bigskip
%\texttt{(iii) The high momentum regime: space-like dispersion relation}
%\bigskip

Consider now further increasing the momentum $\bar k$ of the mode $n$,
beyond the critical value $\bar k_n\approx \sqrt{2} n R_0^3$ at which the
dispersion relation crosses the light--cone. Then the virtuality of the
mode $\bar Q_n$ will increase as well, \ie the mode becomes more and more
space--like (although, as we shall see, this virtuality remains
relatively small, in the sense that $\bar\omega_n\simeq \bar k\gg \bar
Q_n$). For instance, \eqnum{smallk} implies that, so long as $\kappa\ll
\Omega$, the virtuality grows with the energy (or the momentum) according
to
 \beq
 \kappa_n\,\sim\,\Omega_n\left(\frac{\Omega_n-n}{\Omega_n}\right)^{2/3}\,,
 \qquad\mbox{or}\qquad \bar Q_n\,\sim\,\frac{\bar k^{1/3}}{R_0^2}
 \,\Big(\bar k-\bar k_n\Big)^{2/3}\,.\eeq
Eventually, when $\Omega_n\gg n$, $\kappa_n$ becomes comparable with
$\Omega_n$ and then the potential \eqref{Vl} has the shape shown in
Fig.~\ref{fig:Vcd} (for $\ell=0$ and two different values of $\kappa$).

As manifest on these pictures, when increasing the ratio $\kappa/\Omega$,
the turning point $\z_0=1-\kappa/4\Omega$ in the potential moves towards
$\z=0$, \ie towards the bottom of the D7--brane. Thus, clearly, the
attractive region of the potential, which can support Schr\"odinger bound
states with energy $E_n=0$ (or, equivalently, space--like mesons), exists
only so long as $\kappa\le 4\Omega$, and becomes very tiny ($\z_0\ll 1$)
when $\kappa$ approaches the upper limit $4\Omega$ (cf.
Fig.~\ref{fig:Vcd} right). What we would like to argue in what follows is
that for sufficiently large momentum $\bar k\gg nR_0^3$, the dispersion
relation approaches this kinematical limit in which $\kappa_n \simeq
4\Omega_n$: this is the `limiting velocity' regime, previously mentioned
in relation with \eqnum{seer} \cite{mateos,Ejaz:2007hg}.

To that aim, let us compute the spectrum in the regime where $\kappa$ is
indeed close to, but smaller than $4\Omega$, in such a way that $\z_0\ll
1$. The corresponding modes will be localized in the classically
permitted region at $\z\le \z_0$. It is then a good approximation to
replace the potential \eqref{Vl} by its expansion near $\z=0$. The
ensuing Schr\"odinger--like equation reads
 \beq\label{Coulomb}
  \left(-\del_\z^2 \,+\,\frac{\ell(\ell+2)}
 {4\z^2} \,-\,\frac{2e^2}{\z}\,+\,2{\cal E}\right)\psi_l(\z)
 \,=\,0\,,\eeq
where we have denoted
 \beq\label{e2E}
 2e^2&\,\equiv\,&4\Omega^2\,-\,\frac{\kappa^2}{4}\,\simeq\,8\Omega\left(
 \Omega-\frac{\kappa}{4}\right)\,,\nn
 2{\cal
 E}&\,\equiv\,&4\Omega^2\,+\,\frac{\kappa^2}{4}\,\simeq\,8\Omega^2\,.
 \eeq
When deriving \eqnum{Coulomb} and also when simplifying the expressions
in \eqnum{e2E} we have anticipated the fact that, for the mode $n$, both
$\Omega$ and $\kappa$ are very large but relatively close to each other,
such that
 \beq \Omega_n\,,\,\kappa_n\,\gg\,n\,,\qquad \Omega_n\,-\,\frac{\kappa_n}{4}\,
 \simeq\,\frac{n}{\sqrt{2}}\,.\eeq
Also, we restricted ourselves to angular momenta $\ell\ll \Omega_n$.

\eqnum{Coulomb} is formally similar to the radial Schr\"odinger equation
for a non--relativistic particle with mass $m=1$ and electric charge $e$
in the three--dimensional Coulomb potential $V_C(\z)=(-e)/\z$, with
$-{\cal E}$ playing the role of the (negative) energy of a bound state.
There are however some interesting differences with respect to the
genuine Coulomb problem. First, the would--be `angular' momentum of our
fictitious `Coulomb particle' is equal to\footnote{In the usual Coulomb
problem in quantum mechanics, the centrifugal piece of the potential
reads $l(l+1)/\z^2$, which is formally the same as that in
\eqnum{Coulomb} provided one identifies $l=\ell/2$.} $\ell/2$, and hence
it can also take half--integer values. Second, our radial variable $\z$
is restricted to $\z\le 1$ and, moreover, the approximate equation
\eqref{Coulomb} is valid only for $\z\ll 1$; by contrast, in the
corresponding Coulomb problem the radius $\z$ can be arbitrarily large.
Yet, this last difference should not be important for the situation at
hand: given the potential barrier at $\z\le \z_0$, cf.
Fig.~\ref{fig:Vcd}, it is clear that the actual meson wavefunction is
exponentially decaying for $\z > \z_0$ before exactly vanishing at
$\z=1$. When $\z_0\ll 1$, there should be only a minor difference between
the exact wavefunction, which is strictly zero at $\z=1$, and its
Coulombic approximation, which is exponentially small there.

Hence, one can solve \eqnum{Coulomb} by following the same steps as for
the Coulomb problem in quantum mechanics \cite{LLQM}. The general
solution which is regular at $\z=0$ reads\footnote{The notation $z$ for
the rescaled radial coordinate will be used only temporarily and should
not be confused with the respective Minkowski coordinate, which is not
anymore explicit in our analysis.}
 \beq\label{psiCoul}
 \psi_{\ell}(z)&\,=\,&z^{\frac{\ell}{2}+1}\,\rme^{-z/2}\,
 M(-\nu+1+{\ell}/{2}, \ell+2;z)\,,\\*[0.2cm]
 z&\,\equiv\,&2\sqrt{2{\cal E}}\,\z\,,\qquad\nu\,\equiv\,\frac{e^2}
 {\sqrt{2{\cal E}}} \label{nu}\,.\eeq
Here $M(a,b;z)$ is the confluent hypergeometric function, also denoted as
${}_1F_{1}(a,b;z)$, which obeys $M(a,b;0)=1$ (see Chapter 13 in
\cite{AbramStegun}). Note that $z\simeq 4\sqrt{2}\Omega\z$, cf.
\eqnum{e2E}, hence $\z=1$ corresponds to large $z\gg 1$, where the
asymptotic behaviour of \eqnum{psiCoul} becomes relevant. Specifically,
for the solution to exponentially vanish at $z\gg 1$, the confluent
hypergeometric series must terminate, which in turn requires
 \beq\label{nCoul}
 -\nu+1+\frac{\ell}{2}\,=\,-n\,,\qquad n=0,\,1,\,2\,,...
 \eeq
and then $M(-n, \ell+2;z)$ is a polynomial in $z$ of degree $n$
(actually, a Laguerre polynomial $L^{(\ell+1)}_n(z)$, up to a numerical
factor \cite{AbramStegun}). The `quantization' condition \eqref{nCoul}
together with the definitions \eqref{e2E} and \eqref{nu} can now be used
to deduce the meson spectrum in this high momentum regime. The resulting
dispersion relation can be written in various, equivalent, ways, either
as a function of the virtuality $\bar Q$,
 \beq\label{omegaQ}
\bar\omega_{n\ell}(\bar Q)\,-\,\frac{1}{2}\,R_0^2\,\bar
Q\,\simeq\,R_0^3\left(n+{\ell}/{2}+1\right)\,,
\eeq
or as a function of the momentum $\bar k$,
  \beq\label{omegak}
\bar\omega_{n\ell}^2(\bar k)\,-\,v_0^2\bar
k^2\,\simeq\,8\left(n+{\ell}/{2}+1\right)\frac{\bar k}{R_0}\,,
\eeq
or, finally, in physical units:
 \beq\label{omegakphys}
 \omega_{n\ell}^2(k)\,-\,v_0^2
k^2\,\simeq\,(4\pi T)^2\left(n+{\ell}/{2}+1\right)\frac{k}{ M_{\rm gap}
}\,.
\eeq
The last two equations feature the limiting velocity $v_0$ which has been
generated here as
 \beq\label{v0}
 v_0^2\,=\,1\,-\,\frac{4}{R_0^4}\,,\eeq
which is indeed consistent with \eqnum{vm} (recall that we assume $R_0\gg
1$). The results \eqref{omegaQ}--\eqref{omegakphys} are in agreement with
a previous analytic study of this high--momentum regime, in
Ref.~\cite{Ejaz:2007hg}, which is more precise than ours. In any of these
equations, the two terms appearing in the left hand side are large but
comparable with each other, while the term in the right hand side, which
expresses the deviation from the linear dispersion relation $\omega=v_0k$
and involves the dependence upon the quantum numbers, is comparatively
small.

By inspection of the meson wavefunction \eqnum{psiCoul}, where we recall
that $z\simeq 4\sqrt{2}\Omega\z$, it is clear that the mode is localized
near the bottom of the D7--brane, at $\z\lesssim 1/\Omega_n\ll 1$. This
domains lies within the classically allowed region at $\z\le \z_0$
(indeed, $\z_0\equiv 1-\kappa/4\Omega\sim n/\Omega_n$ for a mode
satisfying \eqref{omegaQ}), which confirms the consistency of our
previous approximations. Interestingly, the higher the energy is, the
stronger is the mode localized near $\z=0$ (or $\rho= R_0$), in agreement
with the numerical findings in Ref.~\cite{mateos}.

Although the limiting velocity \eqref{v0} is very close to 1 under the
present assumptions, the virtuality $\bar Q^2=\bar k^2-\bar\omega^2$ of
the meson is nevertheless very large, $\bar Q\gg nR_0$, because its
energy and momentum are even larger: $\bar\omega_n\simeq \bar k\gg
nR_0^3$. Thus the mode looks nearly light--like in the sense that
$\bar\omega_n\simeq \bar k\gg \bar Q$, yet its virtuality is too high to
be resonantly excited by an incoming space--like current: indeed, to be
resonant with the meson, the flavor current should have an energy
$\bar\omega$ and virtuality $\bar Q$ obeying $\bar\omega\simeq
(R_0^2/2)\bar Q \gg nR_0^3$, and therefore $\bar\omega/\bar Q^3 \sim
(R_0/\bar Q)^2 \ll 1$. According to the discussion in
Sect.~\ref{sec:DIS}, such a current would encounter a large repulsive
barrier near the Minkowski boundary and hence it would get stuck at large
radial coordinates $\rho\gtrsim \bar Q\gg nR_0$, far away from the region
at $\rho\simeq R_0$ where the would--be resonant mesons could exist. We
conclude that such high--energy, space--like, meson excitations cannot
contribute to the DIS of a flavor current. To investigate the possibility
of DIS, we therefore turn to the only potentially favorable case, that of
the `nearly light--like' mesons with $\bar \omega_n\simeq \bar k\sim
nR_0^3$ and arbitrarily small virtualities.

\section{Resonant deep inelastic scattering off the light--like mesons}
\setcounter{equation}{0} \label{sec:resDIS}

We now return to the problem of DIS off the strongly coupled ${\cal N}=2$
plasma at low temperature, as formulated in Sect.~\ref{sec:DIS}. Recall
that we are interested in a relatively hard space--like flavor current,
with virtuality $\bar Q> R_0$ (or $\kappa> 1$). So long as the energy
$\bar\omega$ of this current is relatively low, $\bar\omega\ll \bar Q^3$
(or $\Omega\ll \kappa^3$), there is a repulsive barrier which confines
the dual gauge field $A_i(\rho)$ near the Minkowski boundary, where no
resonant meson states can exist. (This barrier is visible in
Fig.~\ref{fig:Vcd} as the repulsive potential at $\z>\z_0$.) We thus
conclude that the flavor structure functions vanish when $\bar\omega\ll
\bar Q^3$, so like for the ${\cal R}$--current.

%because, for that particular kinematics, mesons would be localized near
%the bottom of the D7--brane, at $\rho\sim R_0$.

However, the situation changes when the energy of the current is
sufficiently high, such that $\bar\omega\gtrsim \bar Q^3$ (or
$\Omega\gtrsim \kappa^3$). Then, the repulsive barrier becomes so narrow
that it plays no role anymore (this is visible as the curve
`$\kappa^2>0$' in Fig.~\ref{fig:Vab} right). Indeed, even in the presence
of this barrier, the gauge field can penetrate across the barrier,  via
tunneling, up to a distance $\rho\sim \bar Q$, or $1-\z\sim 1/\kappa^2$~;
when $\Omega\gtrsim \kappa^3$, this penetration is larger then the width
$1-\z_0=\kappa/4\Omega$ of the potential barrier, and then the field can
escape in the classical allowed region at $\z < \z_0$. Thus, the field
has now the capability to excite vector mesons at any value of $\rho$
within the wordvolume of the D7--brane. Moreover the kinematics of the
high--energy current matches with that of the nearly light--like mesons
discussed in Sect.~\ref{sec:LL}. Indeed, those mesons have energies
$\bar\omega_n\sim n R_0^3$, cf. \eqnum{WKBn}, which can match the energy
$\bar\omega\gtrsim \bar Q^3 > R_0^3$ of the current with a suitable
choice for $n$. Furthermore, for a given $n$, there are mesons at all
virtualities $\bar Q_n\lesssim nR_0$, and in particular such that $\bar
Q^3_n\lesssim \bar\omega_n$, so like for the current.

%Interestingly, it is the `high-energy' kinematics of the current which
%matches with the `intermediate energy' kinematics of the mesons.

These kinematical arguments indicate that the high--energy flavor current
can disappear into the plasma by resonantly exciting nearly light--like
mesons. In what follows, we shall demonstrate that this picture is indeed
correct, by explicitly computing the decay rate (\ie the imaginary part
of the current--current correlator) corresponding to the resonant
excitation of large--$n$ light--like mesons. The restriction to large
quantum numbers $n\gg 1$, that is, to very high energies $\bar\omega\gg
R_0^3$, is necessary for technical convenience, but it also has the
advantage to make the physics sharper. In that case, it becomes possible
to smear our the delta--like resonances associated with the individual
mesons and thus obtain a spectral function which is a continuous function
of $\bar\omega$ and hence describes DIS. Remarkably, that spectral
function turns out to be identical with the DIS structure function in the
{\em high--temperature} phase (`black hole embedding'), \eqnum{F1flHT},
that was previously computed \cite{Bayona:2009qe} by imposing infalling
boundary conditions at large $\rho\gg R_0$ (cf. the discussion in
Sect.~\ref{sec:R}).

\subsection{Light--like mesons: exact solutions}
\label{sec:exact}

In this subsection we shall concentrate on the EOM for light--like,
transverse, gauge fields in the worldvolume of the D7--brane, that is,
\eqnum{EAi} with $\bar Q=0$, for which we shall construct exact solutions
obeying the condition of regularity at $r=0$ (or $\rho=R_0$). By using
the asymptotic expansion of these solutions at large $\rho$ and high
energy, we shall study their behaviour near the Minkwoski boundary and
thus distinguish between normalizable and non--normalizable modes. In
particular, this procedure will yield the spectrum of the light--like
mesons for large quantum numbers $n\gg 1$.

%An alternative, approximate, method which relies on the WKB approximation
%and yields the same spectrum at large $n$ will be described in Appendix
%\ref{app:WKB}.

Once again, it is more convenient to use the variable $\z$ defined in
\eqnum{zeta} and which has a compact support. Then the relevant EOM is
\eqnum{EPhi} with $\kappa=0$, or, equivalently, the `Schr\"odinger
equation' \eqref{Epsi} with $\ell=0$ and $\kappa=0$. We shall choose the
latter, that we rewrite here for convenience:
 \beq\label{ELL}
 \left(-\del_\z^2 \,-\,\frac{4\Omega^2(1-\z)}{\z}\right)
 \psi(\z)\,=\,0\,.
 \eeq
Clearly, this is a particular case\footnote{In spite of this formal
similarity, one should keep in mind that \eqnum{Coulomb} and,
respectively, \eqnum{ELL} apply to different physical regimes. In
particular, \eqnum{ELL} holds for any $\z \le 1$, whereas \eqnum{Coulomb}
is valid only for $\z\ll 1$.} of \eqnum{Coulomb} that we have solved
already, namely it is the limit of that equation when $\ell=0$ and
${2e^2}=2{\cal E}\equiv 4\Omega^2$. The solution which is regular at
$\z=0$ is then obtained by adapting \eqnum{psiCoul}, and reads
\beq\label{psiLL}
 \Phi(\z)\,\equiv\,\frac{1}{\z}\,\psi(\z)\,=\,\rme^{-2\Omega \z}\,
 M(-\Omega+1, 2; 4\Omega\z)\,,\eeq
which once again features the confluent hypergeometric function
$M(a,b;z)$. This solution takes on a finite value on the Minkowski
boundary at $\z=1$ and in general, \ie for generic values of the energy
parameter $\Omega$, it represents a non--normalizable mode. The
normalizable modes describing mesons are obtained by requiring the
solution to vanish at $\z=1$:
 \beq\label{LLn}
 M(-\Omega_n+1, 2; 4\Omega_n)\,=\,0,\eeq
with $\Omega_n$ the on--shell energy of a light--like meson. For generic
values $\Omega\sim \order{1}$, this equation is difficult to solve except
through numerical methods. A similar mathematical difficulty arises when
trying to use \eqnum{psiLL} in order to compute the current--current
correlator according to \eqnum{Pi1}. For all such purposes, one needs the
behavior of the solution near the Minkowski boundary at $\z=1$, and this
is generally difficult to extract from \eqnum{psiLL}.

However this mathematical problem becomes tractable for the high energy
regime of interest here, which is such that $\bar\omega\gg R_0^3$, or
$\Omega\gg 1$. Then, one can use a special asymptotic expansion of the
function $M(a,b;z)$ with $a<0$, which applies when the variables $|a|$
and $z$ are simultaneously large and such that $z\approx 2b-4a\gg 1$.
This last condition is truly essential, since in general, \ie for generic
values of $|a|$ and $z$ which are both large but uncorrelated with each
other, very little is known about the asymptotic behavior of $M(a,b;z)$.
This specific limit is precisely the one that we need for our present
purposes: indeed, in \eqnum{psiLL}, we have $z=4\Omega\z$ and
$2b-4a=4\Omega$, and therefore $z\sim 2b-4a=4\Omega\gg 1$ in the high
energy limit and in the vicinity of $\z= 1$. The asymptotic formula which
applies to this case is formula 13.5.19 in Ref.~\cite{AbramStegun} and
can be formulated as follows: when
 \beq\label{xi}
 z\,=\,(2b-4a)\left[1+\,\frac{\xi}{(b-2a)^{{2}/{3}}}\right]
 \quad\mbox{with \ \ $|\xi|\,\ll\,(b-2a)^{{2}/{3}}$}\,,
 \eeq
then (below, $\Gamma(b)$ is the Euler function, and Ai and Bi are the
Airy functions)
 \beq\label{ASasymp}
 M(a,b;z)\,=\,\rme^{\frac{z}{2}}\,\big(b-2a\big)^{\frac{2}{3}-b}\,\Gamma(b)
 \left[{\rm Ai}(\xi)\cos(\pi a)+{\rm Bi}(\xi)\sin(\pi a) \,+\,
 \order{|b-2a|^{-1}}\right]\,.
 \eeq
In order to adapt this formula to \eqnum{psiLL}, we shall write
 \beq\label{xiOmega}
 \z\,\equiv\,1\,-\,\frac{\xi}{(2\Omega)^{{2}/{3}}}\,,\eeq
so that the variable $\xi$ be positive. Then for $\Omega\gg 1$ and
$1-\z\ll 1$, the solution \eqref{psiLL} becomes (up to an irrelevant
overall normalization)
 \beq\label{psiLLas}
 \Phi(\xi)\,\simeq\,{\rm Ai}(-\xi)\cos(\Omega \pi)-{\rm Bi}(-\xi)
 \sin(\Omega\pi)\qquad\mbox{for \ \ $0\,\le\, \xi\,\ll\,(2\Omega)^{{2}/{3}}$}
 \,.\eeq
In particular, for relatively large $\xi\gg 1$, one can use the
asymptotic expansions of the Airy functions to deduce
  \beq\label{psiLLinter}
 \Phi(\xi)\,\simeq\,
 \frac{1}{\sqrt{\pi}\xi^{1/4}}
 \,\sin\left(\frac{2}{3}\,\xi^{3/2}\,+\,\frac{\pi}{4}\,-\,\Omega{\pi}
 \right)\quad\mbox{for}\quad 1\,\ll\,\xi\,\ll\,(2\Omega)^{{2}/{3}}
 \,.\eeq

The solution  \eqref{psiLLas} has the same general structure and validity
range as the approximate solution shown in \eqnum{AiAiry} --- in
particular, the argument $\xi$ of the Airy functions is indeed the same
in both equations, as it can be checked by using Eqs.~\eqref{zeta} and
\eqref{xiOmega} ---, and this should not be a surprise: as explained in
Sect.~\ref{sec:DIS}, \eqnum{AiAiry} is the general form of the solution
at high energy and large $\rho$. The whole purpose of a more complete
analysis at smaller values of $\rho$, like the one that we have just
performed here, is to fix the coefficients of the two Airy functions
appearing in that equation. Note that, unlike for the solution with
infalling boundary condition,  \ie \eqnum{AiAiry} with $c=i$, the
coefficients in \eqnum{psiLLas} depend upon the energy variable $\Omega$.

% (compare in that respect \eqnum{psiLLinter} to \eqnum{AiRsol}).

As a first application of \eqnum{psiLLas}, we now use it to determine the
energies of the light--like meson excitations according to \eqnum{LLn}.
To that aim, we also need \cite{AbramStegun}
 \beq\label{Airy0}
 {\rm Ai}(0)\,=\,\frac{1}{3^{2/3}\Gamma(2/3)}
 \,=\,\frac{1}{\sqrt{3}}\,{\rm Bi}(0)\,,\qquad
 {\rm Ai}'(0)\,=\,-\frac{1}{3^{1/3}\Gamma(1/3)}
 \,=\,-\frac{1}{\sqrt{3}}\,{\rm Bi}'(0)\,,\eeq
(the formul\ae{} involving the derivatives will be useful later on). Then
a simple calculation shows that for $\xi=0$ the right hand side of
\eqnum{psiLLas} is proportional to $\sin[(\Omega-1/6)\pi]$, and then
\eqnum{LLn} immediately implies
  \beq\label{Omegan}
 \Omega_n\,=\,n\,+\,\frac{1}{6}\,,\eeq
in agreement with \eqnum{WKBn}. It can be numerically checked that
\eqnum{Omegan} is a very good approximation to the zeroes of \eqnum{LLn}
already for small values $n=1,2,...$. One may think that the constant
shift $\Omega_n-n=1/6$ in the eigenvalues is merely a tiny correction
that can be safely ignored at large $n$, but this is generally not the
case. Note first that this shift is uniquely determined by the values of
the two Airy functions at $\xi=0$, as it can be checked by inspection of
the previous manipulations, and hence it is not affected by the
approximation in \eqnum{psiLLas}. Moreover it is essential to take this
shift into account whenever one is interested in the behavior of the
solution near $\Omega=\Omega_n$, which will be also our case in the next
subsection.

The wavefunction corresponding to the mode $n$ is obtained by replacing
$\Omega\to\Omega_n$ within the general formul\ae{} \eqref{psiLL} or
\eqref{psiLLas}. For radial coordinates deeply inside the D7--brane,
where the asymptotic expansion \eqref{psiLLas} does not apply, one can
rely on the exact solution \eqref{psiLL}, but this is perhaps a little
opaque. An approximate formula valid for intermediate values of $\z$ will
be constructed via the WKB method in Appendix \ref{app:WKB}. This WKB
solution, shown in \eqnum{WKBpsi1}, is consistent with the asymptotic
behaviour \eqref{psiLLinter} and has the nice feature to exhibit exactly
$n$ nodes in the interval $0< \z < 1$, as {\em a priori} expected for the
$n$th radial excitation.

 \comment{
Near the boundary, the corresponding wavefunction is given by
  \beq\label{psiLLn}
 \Phi_n(\xi)\,\simeq\,(-)^n
 \left[ \frac{\sqrt{3}}{2}\,
 {\rm Ai}(-\xi_n)\,-\, \frac{1}{2}\,
 {\rm Bi}(-\xi_n)\right]
 \qquad\mbox{for \ \ $0\,\le\, \xi\,\ll\,(2\Omega_n)^{{2}/{3}}$}
 \,,\eeq
with $\xi_n$ defined by \eqnum{xiOmega} with $\Omega\to\Omega_n$.

In particular, for relatively large $\xi\gg 1$, one can use the
asymptotic expansions of the Airy functions to deduce
  \beq\label{psiLLnas}
 \Phi_n(\xi)\,\simeq\,
 \frac{(-)^n}{\sqrt{\pi}\xi_n^{1/4}}
 \,\sin\left(\frac{2}{3}\,\xi_n^{3/2}+ \frac{\pi}{12}
 \right)\quad\mbox{for}\quad 1\,\ll\,\xi\,\ll\,(2\Omega_n)^{{2}/{3}}
 \,,\eeq
where $\frac{\pi}{12}$ has been generated as
$\frac{\pi}{12}=\frac{\pi}{4}-\frac{\pi}{6}$. It is interesting to
compare this to \eqref{AiRsol}.}

\subsection{Current--current correlator and DIS}

We are now prepared to compute the current--current correlator for a
highly energetic, nearly light--like, current, and thus make the
connection to DIS, as anticipated. To that aim, we shall use \eqnum{Pi1}
together with the asymptotic expansion of the solution near the Minkowski
boundary, \eqnum{psiLLas}. Specifically, \eqnum{Pi1} involves (recall
that $A_i\equiv\Phi$ in our present notations, and the radial coordinate
$r$ in \eqnum{Pi1} is related to the variable $\xi$ which appears in
\eqnum{psiLLas} via Eqs.~\eqref{zeta} and \eqref{xiOmega})
 \beq
\left[r^3\,\frac
 {\del_r A_i(r,\omega,k)}{A_i(r,\omega,k)}\right]_{r\to\infty}\,=\,
 -4\left(\frac{\bar\omega}{2}
   \right)^{{2}/{3}}\left[
   \frac{\del_\xi \Phi(\xi,\Omega)}{\Phi(\xi,\Omega)}
  \right]_{\xi\to 0}\,.\eeq
Then a straightforward calculation using $\Phi(\xi,\Omega)$ from
\eqref{psiLLas} together with \eqnum{Airy0} yields
 \beq\label{Piflav}
  \Pi_1(\bar\omega)\,=\,-
  \frac{3\pi\nc\nf T^2}{2\Gamma^2(1/3)}\left(\frac{\bar\omega}{6}
   \right)^{{2}/{3}}\left[
   \frac{1}{\sqrt{3}}\,+\,\cot\left(\Omega-\frac{1}{6}\right)\pi
   \right]
   \,,\eeq
with $\Omega$ related to $\bar\omega$ via \eqnum{kapomeg}. As expected,
the function $\Pi_1(\bar\omega)$ exhibits poles at the energies $\Omega_n
= n+1/6$ of the light--like meson modes. These poles can be made more
explicit by using the expansion of the cotangent as a series of simple
functions:
 \beq
 \cot \pi x\,=\,\frac{1}{\pi x}\,+\,\frac{2x}{\pi}
 \sum_{n=1}^\infty \,\frac{1}{x^2-n^2}\,.\eeq
To extract the spectral weight associated with these poles, \ie the
imaginary part of the correlator, we use retarded boundary conditions,
$\omega\to\omega + i\epsilon$, together with the formula
 \beq
 \lim_{\epsilon\to 0}\,\frac{1}{x+i\epsilon -n}\,=\,
 {\cal P}\,\frac{1}{x} \,-\,i\pi\delta(x-n)\,.\eeq
We thus find (recall that our energy variable is always positive)
 \beq\label{ImPiRes}
 {\rm Im}\,\Pi_1(\bar\omega+i\epsilon)\,=\,
 \frac{3\pi}{2\Gamma^2(1/3)}\,\nc\nf T^2
 \left(\frac{\bar\omega}{6}
   \right)^{{2}/{3}}\sum_{n=1}^\infty \,\delta(\Omega-\Omega_n)\,.
   \eeq
This result has been obtained here by working with a light--like current,
but a similar result holds also for a highly energetic space--like
current with $\bar\omega\gtrsim \bar Q^3$, since the repulsive barrier
plays no role in that case, and since the plasma can indeed sustain
slightly space--like mesons which are resonant with the current. (In
fact, the WKB method in  Appendix \ref{app:WKB} can be easily generalized
to such slightly space--like mesons.) The emergence of the
delta--functions in the imaginary part of the current--current correlator
confirms our expectation that a high--energy flavor current can
resonantly produce mesons in highly excited states ($n\gg 1$) and thus
disappear into the plasma.

Taken literally, \eqnum{ImPiRes} would imply that the meson production by
the current can only occur for a discrete set of energies which are
resonant with the energies $\bar\omega_n=\sqrt{2}(n+\frac{1}{6})R_0^3$ of
the light--like meson excitations in the plasma. However, an argument
based on the uncertainty principle together with the peculiar structure
of the meson dispersion relation near the light--cone (cf.
Sect.~\ref{sec:LL}) shows that in order for the absorbtion process to be
observable, one needs to average \eqnum{ImPiRes} over neighboring levels.

The argument goes as follows: A current with a given, large, momentum $k$
which is produced by a source acting over a finite time interval $\delta
t$ has an uncertainty $\delta\omega\sim 1/\delta t$ in its energy, and
hence an uncertainty $\delta Q^2\simeq 2k \delta\omega$ in its
virtuality. (We have used here $k=\sqrt{\omega^2+Q^2}\simeq
\omega+Q^2/2\omega$ at high energy $k\gg Q$.) As we shall demonstrate in
Sect.~\ref{sec:TIME}, via an analysis of the time scales for the current
interactions in the plasma, the typical interaction time for a nearly
light--like current scales with its momentum like
 \beq\label{deltat}
 \bar t_{\rm int}\,\sim\,\bar
 k^{1/3}\,.
 \eeq
(As usual, a bar over a kinematic variable denotes the dimensionless
version of that variable measured in units of $\pi T$, \eg{} $\bar t=\pi
Tt$.) So, for this process to be experimentally observable, the source
producing the current must act over a comparatively short period of time:
$\delta t\lesssim t_{\rm int}$. (If $\delta t \gg t_{\rm int}$, one
cannot distinguish between the absorbtion of the photons in the plasma
and their reabsorbtion by the source.) This in turn implies
 \beq\label{deltaQ}
 \delta\omega\,\gtrsim\,\frac{1}{\bar t_{\rm int}}\,\sim\,\frac{1}{\bar
 k^{1/3}}\,,\qquad\mbox{or}\qquad \delta \bar Q\,\gtrsim\,\bar
 k^{1/3}\,.\eeq
This means that the high--energy flavor current has the potential to
produce meson excitations with momentum $\bar k$ (the momentum of the
current) and virtualities within a range  $\delta \bar Q\gtrsim \bar
k^{1/3}$ around $\bar Q=0$.

At this point one should remember the discussion towards the end of
Sect.~\ref{sec:LL}, about the high sensitivity of the nearly light--like
meson dispersion relation to changes in the virtuality. Let us rephrase
that discussion but from a different perspective: assume that the
momentum $\bar k$ of the meson is now fixed, but consider changes in the
mode quantum number $n$ associated with changes in virtuality $\kappa$
(near $\kappa=0$). As it should be clear from \eqnum{smallk}, $n$ varies
by a number of order one when $\kappa$ changes by $\Delta\kappa
\sim\Omega^{1/3}$, that is, when $\bar Q$ varies by $\Delta \bar Q\sim
\bar k^{1/3}$. This is of the same order as the lower limit on the
uncertainty \eqref{deltaQ} in the virtuality of the current. Thus, for a
given momentum $\bar k$, the current has the possibility to be resonant
with several meson states, with neighboring quantum numbers.

Thus, in order to compute the total interaction rate for the current, as
given by the imaginary part of the current--current correlator, we are
allowed to average \eqnum{ImPiRes} over several neighboring levels and
thus smear our the delta--function resonances. This averaging amounts to
integrating \eqnum{ImPiRes} over an interval $\delta\Omega$ which
contains a few neighboring resonances and dividing the result by
$\delta\Omega$, thus yielding the following result
 \beq\label{ImPi}
 {\rm Im}\,\Pi_1(\bar k)\,=\,
 \frac{3\pi}{2\Gamma^2(1/3)}\,\nc\nf T^2
 \left(\frac{\bar k}{6}
   \right)^{{2}/{3}}\,,
   \eeq
for the imaginary part of the retarded 2--point function of a high energy
flavor current with momentum $\bar k$ and energy $\bar\omega\simeq\bar
k$. In particular, when the current is space--like (with relatively small
virtuality, though: $\bar Q\lesssim \bar k^{1/3}$), a non--zero imaginary
part is synonymous of deep inelastic scattering, and the above result can
be identified with the DIS structure function: $F_1 =(1/2\pi){\rm
Im}\,\Pi_1$. By comparing this result to \eqnum{ImPiRes} we see that the
structure function in this low temperature phase is exactly the same as
in the high--temperature phase, or `black hole embedding'. This
coincidence reflects that fact that in both cases the current is
completely absorbed into the plasma, although the respective mechanisms
are quite different: resonant excitation of nearly light--like mesons at
low temperature and, respectively, partonic fluctuations (a
quark--antiquark pair together with arbitrary many ${\cal N}=4$ quanta),
which disappear into the plasma via successive branching, at high
temperature. This physical picture will be further clarified by a
discussion of the relevant time scales in the next section.

Although our present calculations apply to the transverse field and
structure function alone (recall that $F_T=2xF_1$), it is clear that a
similar argument must be valid also in the longitudinal sector. So the
corresponding, flavor, structure function $F_L=F_2-F_T$ can be deduced by
simply rescaling, by a factor $4\nf/\nc$, the respective result for the
${\cal R}$--current \cite{HIM2}
 \beq\label{FLfl}
  F_L&\equiv &F_2-2xF_1=\frac{\nc\nf Q^2x}{24\pi^2}\,.
  \eeq
To summarize, for the ${\cal N}=2$ plasma at strong coupling, the above
results for the flavor DIS structure functions are valid at either low,
or high, temperatures, for high enough momenta $\bar k\gg R_0^3$ and for
sufficiently low virtualities $\bar Q\lesssim \bar k^{1/3}$. In physical
units, these conditions amount to $k\gg T(M_{\rm gap}/ T)^3$ and
$Q\lesssim Q_s(x)$, where the saturation momentum $Q_s(x)\simeq T/x$ is
the same as for the ${\cal R}$--current.

\section{Time dependence and physical picture}
\setcounter{equation}{0} \label{sec:TIME}

In what follows we would like to provide a space--time picture for the
interactions of the flavor current in the strongly--coupled ${\cal N}=2$
plasma, and thus in particular clarify the energy averaging over
neighboring resonances performed in the previous section. To that aim, we
need to assume that the source producing the current has acted over a
finite interval of time $\delta t$, which is much shorter than the
typical interaction time in the plasma, $t_{\rm int}$, that we shall
compute. Accordingly, the current is not a simple plane--wave anymore,
but rather a wave--packet in energy. To study the dynamics of this
wave--packet, it is convenient to first reformulate the respective EOM as
a time--dependent Schr\"odinger equation. This will also give us insight
into the typical time scales for meson excitations in the plasma.

A current produced over a finite period of time $\delta t$ can be
described as a wave packet in energy, with a width $\delta \omega\sim
1/\delta t$ which is much smaller than the central value $\omega_0\simeq
k$. (We assume that the current has a sharply defined longitudinal
momentum $k$ and we consider the high energy kinematics where
$\omega\simeq k$.) To study the evolution of this wave packet with time,
we need to restore the time--dependence in the respective equations of
motion, as written down in Sect.~\ref{sec:EOM}. For the problem at hand,
this can be readily done by replacing
 \beq
 \omega^2\ \to \ (\omega_0 + i\del_t)^2\, \simeq\, \omega_0^2 +
 2i\omega_0\del_t\,,
 \eeq
in equations like \eqref{EAi}. Indeed, for the (relatively narrow) wave
packet under consideration, the time dependence will be represented by a
wave $\rme^{-i\omega_0 t}$ modulated by a relatively slowly varying
function ($|i\del_t|\ll\omega_0$). Starting with \eqref{EAi}, we thus
obtain
\beq\label{EAit}
  -\frac{4i\bar k}{\rho^4}\,\del_{\bar t} A_i
  \,=\,\ddot A_i +
 \frac{3}{r}\,\dot A_i \,+
 \left(-\frac{2{\bar Q}^2}{\rho^4}
 \,+\,\frac{8\bar k^2}{\rho^8}\right)A_i\,,
 \eeq
where we recall that $\bar t=\pi Tt$, $\bar k=k/\pi T$, $\rho^2
=R_0^2+r^2$, and a dot denotes a derivative w.r.t. $r$. In writing this
equation, we have replaced $\omega_0$ by $k$ everywhere except in the
virtuality term $\bar Q^2=\bar k-\bar \omega_0^2$ (which in what follows
will be treated as a small correction). Also, we have neglected a
subleading term involving the time derivative which is proportional to
$1/\rho^8$ (recall that $\rho\ge R_0\gg 1$). We would like to rewrite
this equation as a time--dependent Schr\"odinger equation, since then we
can rely on the techniques and intuition developed within quantum
mechanics. The first step is to eliminate the term $\propto \dot A_i$, by
writing $A_i(\bar t,r)\equiv \Phi(\bar t,r)/r^{3/2}$ :
 \beq\label{EPsit}
  \frac{4i\bar k}{\rho^4}\,\frac{\del\Phi}{\del \bar t}
  \,=\,\left(-\,\frac{\del^2}{\del r^2}\,+\,\frac{3}{4r^2}
  \,+\,\frac{2{\bar Q}^2}{\rho^4}
  -\,\frac{8\bar k^2}{\rho^8}\right)\Phi\,.\eeq
%where we have limited ourselves to the nearly light--like case $\bar
%Q_0\simeq 0$ (more precisely, $\kappa_0\ll\Omega$, with obvious
%notations), so like in Sect.~\ref{sec:resDIS}.
Given the $r$--dependent factor $1/\rho^4$ multiplying the time
derivative in the l.h.s., this equation is not yet in Schr\"odinger form.
It turns out that the canonical, Schr\"odinger, form of the equation can
be achieved by changing the radial coordinate one more time, namely, by
using the angle $\theta$ introduced in Sect.~\ref{sec:D3D7}, cf.
\eqnum{coord2}, to that purpose. Specifically, after writing
 \beq\label{tau}
  \bar t\,\equiv\,2\sqrt{2}R_0\,\tau\,,\qquad
 \rho\,\equiv\,\frac{R_0}{\sin\theta}\,,\qquad
 \Phi(\bar t,r)\,\equiv\,\frac{1}{\sin\theta}\,\Psi(\bar t,\theta)
 \,,\eeq
one finds that \eqnum{EPsit} takes the canonical form (with $\Omega$ and
$\kappa$ defined as in \eqnum{kapomeg})
 \beq\label{HPsit}
 i\,\frac{\del\Psi}{\del \tau}
  &\,=\,&\left(-\,\frac{1}{2\Omega}\,
  \frac{\del^2}{\del \theta^2}\,+\,V(\theta)\right)\Psi,
  \\*[0.2cm]
  V(\theta)&\,=\,&-\frac{1}{2\Omega}\,-\,
  {8\Omega\sin^4\theta}\,+\,
  \frac{3}{8\Omega}\,\frac{1}{\sin^2\theta\cos^2\theta}\,+\,
  \frac{\kappa^2}{2\Omega}\,.\label{Vtheta}
  \eeq
Formally, these equations describe the quantum dynamics of a
non--relativistic particle with mass $\Omega$ and Hamiltonian $H$ defined
by the r.h.s. of \eqnum{HPsit}.

Note that the term involving the virtuality $\kappa$ within the potential
is independent of $\theta$ and thus merely acts as a constant shift in
the total energy, in the same way as the time derivative. This is as
expected: the time--dependence in the problem arises because of the
uncertainty $\delta\omega$ in the energy of the wave--packet; for a fixed
momentum $k$, this corresponds to an uncertainty $\delta Q$ in the
virtuality, such that $\delta\omega\simeq \delta Q^2/2k$. In the units of
\eqnum{HPsit}, this amounts to $i\del_\tau\sim \delta
{\kappa^2}/{2\Omega}$, so the time--derivative and the (central value of
the) virtuality act indeed on the same footing. This being said, in what
follows we shall often ignore the last term ${\kappa^2}/{2\Omega}$ in
\eqnum{Vtheta}, precisely because we are interested in situations where
the fluctuations in virtuality associated with the uncertainty principle
are larger than the central value $Q^2$. That is, we shall assume $\bar
Q\ll \bar k^{1/3}$ (the high--energy kinematics where the DIS process
becomes possible), whereas we shall see that $\delta \bar Q\gtrsim \bar
k^{1/3}$.

The potential in \eqnum{Vtheta} is displayed in Fig.~\ref{fig:theta} for
the interesting case $\Omega\gg 1$ and $\kappa=0$. The Minkowski boundary
lies at $\theta=0$ and the bottom of the D7--brane at $\theta=\pi/2$.
Other remarkable points are the two classical turning points, $\theta_1$
and $\theta_3$, and the minimum of the potential at $\theta_2$. One finds
 \beq\label{thetai}
 \theta_1\,\simeq\,\frac{3^{1/6}}{2\Omega^{1/3}}\,,
 \qquad
 \frac{\pi}{2}\,-\,\theta_2\,\simeq\,%\left(\frac{3}{2}\right)^{1/4}
 \frac{({3}/{2})^{1/4}}{(8\Omega)^{1/2}}\,,
 \qquad
 \frac{\pi}{2}\,-\,\theta_3\,\simeq\,\frac{3^{1/2}}{8\Omega}\,,
 \eeq
so that $\theta_1\ll 1$, while $\theta_2$ and $\theta_3$ are close to
$\pi/2$. The potential has a rather deep minimum: $V(\theta_2)\simeq
-8\Omega$.

\FIGURE[t]{
\centerline{%\hspace*{-3.cm}
\includegraphics[width=0.8\textwidth]{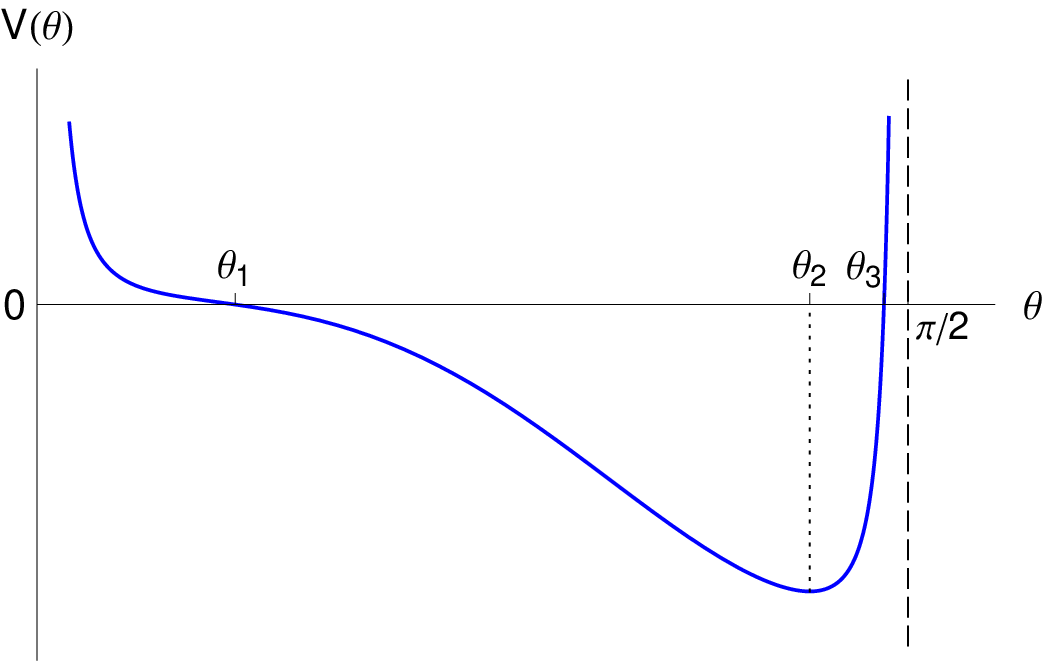}
} \caption{\sl The potential $V(\theta)$ in \eqnum{Vtheta} for $\Omega=5$
and $\kappa=0$. \label{fig:theta}} }

To estimate the interaction time for the flavor current, we shall
evaluate the time that the wave packet takes to travel from the boundary
to the interior of the D7--brane, where it can excite mesons. A similar
analysis for the case of the ${\cal R}$--current was presented in
Ref.~\cite{HIM3}. Near the boundary, where $\theta\ll 1$, the equation
becomes
 \beq\label{HPsibdry}
 i\,\frac{\del\Psi}{\del \tau}
  \,\simeq\,\left(-\,\frac{1}{2\Omega}\,
  \frac{\del^2}{\del \theta^2}\,+\,\frac{3}{8\Omega}
  \frac{1}{\theta^2}\right)\Psi,
  \eeq
with the following, exact, solution ($C$ is a constant)
 \beq\label{diff}
 \Psi(\tau,\theta)\,=\,C\,\frac{\theta^{3/2}}{\tau^2}\
 \exp\left(
 i\frac{\Omega\theta^2}{2\tau}\right)\,,\eeq
which describes {\em diffusion}\,: at early times, the penetration
$\theta$ of the wave packet in the radial dimension grows with $\tau$
like
 \beq
 \theta(\tau) %\equiv \frac{\pi}{2}-\theta(\tau)
 \,\sim\,\sqrt{\frac{2\tau}{\Omega}}
 \qquad\mbox{for}\quad \tau\,\lesssim\,\tau_1\,\sim\,\Omega^{1/3}\,.
 \eeq
Here $\tau_1$ is the time the wave takes to reach the point $\theta_1$,
cf. \eqnum{thetai}, where the potential becomes attractive; in physical
units, $\bar t_1\sim R_0\tau_1 \sim \bar k^{1/3}$.

For $\tau>\tau_1$, and so long as $\theta$ lies in between $\theta_1$ and
the minimum of the potential at $\theta_2$, the wave packet falls in the
potential, essentially by following the classical equation of motion:
 \beq
 \Omega \,\frac{\del^2 \theta}{\del\tau^2}\,=\,-\,\frac{\del V}
 {\del \theta}\,\simeq\,8\Omega\,\frac{\del}
 {\del \theta}\,\sin^4\theta\,.\eeq
For most of this travel, $\theta$ is still small, so $\sin^4\theta\simeq
\theta^4$ and
  \beq
\frac{\del^2 \theta}{\del\tau^2}\,\simeq\,32\, \theta^3\ \Longrightarrow\
 \theta(\tau)\,\simeq\,\frac{1}{1/\theta_1 - 4(\tau-\tau_1)}\,.\eeq
So, the penetration $\theta$ becomes of $\order{1}$ at $\tau\sim\tau_2$,
with
 \beq
 \tau_2-\tau_1\, \sim\, \frac{1}{4\theta_1} \,\sim\, \Omega^{1/3}
 \ \Longrightarrow\ \bar t_2 - \bar t_1 \,\sim\,\bar t_1\,\sim\,
 \bar k^{1/3}\,.\eeq
That is, it takes (parametrically) as much time to the wave to fall in
the potential up to large distances $\theta\sim \order{1}$ as it takes to
diffusively reach the attractive part of the potential, although the
respective distances are very different: $\theta_2\gg\theta_1$. This is
so because along the second part of the trajectory, from $\theta_1$ to
$\theta_2$, the wave has un accelerated motion under the influence of the
BH.

This time $\bar t_2\sim \bar k^{1/3}$ is already a realistic estimate for
the interaction time, since the current can resonantly produce mesons
when $\theta$ is of $\order{1}$. Moreover, this estimate would not change
if the meson was to be produced further down in the worldvolume of the
D7--brane (say, near its lower tip at $\theta\simeq \pi/2$), since the
final part of the fall in the potential is very rapid because the
potential is so attractive around $\theta_2$. We thus conclude that $\bar
t_{\rm int}\sim \bar k^{1/3}$, as anticipated in \eqnum{deltat}. This
also confirms that the typical uncertainties in the energy and the
virtuality of the wave packet satisfy $\delta\bar\omega\gtrsim 1/\bar
k^{1/3}$ and, respectively, $\delta \bar Q\gtrsim \bar k^{1/3}$, in
agreement with our original assumptions.

Since determined by the dynamics relatively close to the Minkowski
boundary, this interaction time is independent of $R_0$, and thus is
parametrically the same as for the ${\cal R}$--current \cite{HIM3}. As we
shall shortly see, a similar conclusion holds also for the light--like
meson excitations with large quantum numbers $n\gg 1$ : such a meson has
a period (the interval of time corresponding to one rotation around a
semiclassical orbit) $\Delta \bar t_n \sim \bar\omega^{1/3}_n$ and spends
most of this time at radial locations far away from $R_0$, namely around
$\rho\sim n^{1/3}R_0$. The subsequent calculation will also shed light on
an interesting property of the meson spectrum, which played an important
role in our previous argument: the strong sensitivity of the dispersion
relation to variations in the virtuality $\bar Q$ around the light--cone
($\bar Q=0$).

To that aim, we resort again on the Bohr--Sommerfeld quantization
formula, valid for large $n$. We thus write
 \beq\label{BS2}
 n\pi\,=\,\int_{\theta_1}^{\theta_3}
 \rmd\theta\,p(\theta\,; E_n,\Omega)\,,\qquad
 p(\theta\,; E,\Omega)\,\equiv\,\sqrt{2\Omega(E - V(\Omega))}\,,\eeq
where $E_n$ are the energy levels associated with the Schr\"odinger
Hamiltonian in \eqnum{HPsit}. These energies are defined by the usual
eigenvalue problem $H\Psi_n=E_n\Psi_n$ and depend upon the two parameters
$\Omega$ and $\kappa$ within $H$. They should not be confused with the
meson energies $\Omega_n$, which rather correspond to the special values
of $\Omega$ (at a given $\kappa$) for which the homogeneous equation
$H\Psi=0$ has non--trivial, normalizable, solutions. One clearly has
 \beq
 E_n(\Omega,\kappa)\,=\,0\qquad\mbox{for}\qquad \Omega=
 \Omega_n(\kappa)\,.\eeq
This last equation has a unique solution for any $n=0,1,2,...$, as it can
be easily recognized by inspection of the potential in \eqnum{Vtheta} and
Fig.~\ref{fig:theta}. By choosing $E_n=0$ in \eqnum{BS2}, one could work
out the corresponding integral and thus recover our previous result that,
\eg, $\Omega_n(\kappa)=n$, cf. \eqnum{WKBn}. However, for the present
purposes, we shall rather use \eqnum{BS2} to compute the {\em level
spacing} of the Schr\"odinger energies near $E=0$, that is
 \beq
 \Delta E(\Omega_n)\,\equiv\,E_{n+1}(\Omega_n)-E_n(\Omega_n)
 \,=\,E_n(\Omega_n)\qquad (\kappa\,=\,0)\,.\eeq
Using \eqnum{BS2}, this is obtained as (we anticipate that $\Delta E\ll
1$)
 \beq\label{deltaE}
 \pi\,=\,\int_{\theta_1}^{\theta_3}\rmd\theta\,\big[
 p(\theta\,; \Delta E,\Omega_n)-
 p(\theta\,; 0,\Omega_n)\big]
 \,\simeq\,\Delta E \int_{\theta_1}^{\theta_3}\rmd\theta\,\frac
 {\del p}{\del E}\Bigg|_{E=0}\,=\,\Delta E
 \int_{\theta_1}^{\theta_3}\rmd\theta\,\frac
 {\Omega_n}{p}\,.
 \eeq
Indeed, this quantity $\Delta E(\Omega_n)$ provides the answer to the two
questions that we are interested in, as we explain now:

\bigskip
\texttt{(i)} The last integral in \eqnum{deltaE} is the same as half of
the period for a round trip around a semiclassical orbit: indeed,
$p=\Omega\del_\tau\theta$, hence $(\Omega/p) \rmd\theta=\rmd\tau$.

\bigskip
\texttt{(ii)} The quantity $\Delta E(\Omega_n)$ characterizes the
response of the meson dispersion relation to variations in the meson
virtuality near the light--cone. This can be understood by recalling the
discussion below \eqnum{Vtheta}, about the last term,
${\kappa^2}/{2\Omega}$, in the potential. More precisely, $\Delta
E(\Omega_n)$ is a measure of the increase in the virtuality which is
needed to jump from one meson level ($n+1$) to the neighboring one ($n$)
at fixed energy and in the vicinity of the light--cone. In formul\ae, $
\Delta E(\Omega_n)={\kappa^2_n}/{2\Omega_n}$ with $\kappa_n$ defined by
$\Omega_{n+1}(0)=\Omega_n(\kappa_n)$, or
 \beq
 E_n(\Omega_n(\kappa_n),\kappa_n)\,=\,E_{n+1}(\Omega_n(\kappa_n),0)
 \,=\,0\,.\eeq

Returning to \eqnum{deltaE}, this can be evaluated as
 \beq\label{period}
 \frac{\pi}{\Delta E}
 &\,\simeq\,&\frac{1}{4}
 \int_{\theta_1}^{\theta_3}\rmd\theta\,\left(
 {\sin^4\theta \,-\,
 \frac{3}{64\Omega_n^2}\,\frac{1}{\sin^2\theta\cos^2\theta}}\right)
 ^{-1/2}\\*[.2cm]\nonumber
 &\,\simeq\,&\frac{1}{4}
 \int_{\theta_1}^{\infty}\,\frac{\rmd\theta \,\theta}
 {\sqrt{\theta^6 - \theta^6_1}}\,=\,\frac{1}{24\theta_1}\,{\rm
 B}(1/6,1/2)\,,
 \eeq
where in the second line we have used the fact that the integral  is
dominated by its lower limit $\theta_1\ll 1$, and ${\rm B}(1/6,1/2)$ is
the respective Beta function.

We thus find that the period for motion of a meson around the
semiclassical orbit with quantum number $n$ is $\Delta \bar t_n \sim
1/\theta_1 \sim \bar\omega^{1/3}_n$. Moreover, the fact that the above
integral is dominated by $\theta\sim \theta_1$ also means that a quantum
particle in the bound state with energy $E_n\simeq 0$ spends most of its
time at relatively large radial distances $\rho\sim\bar\omega^{1/3}_n\sim
n^{1/3}R_0$. This implies a similar property for the light--like meson
with energy $\bar\omega_n\simeq\sqrt{2}nR_0^3$. Of course, the meson
wavefunction has support everywhere in the range $R_0\lesssim \rho
\lesssim n^{1/3}R_0$, but the radial velocity $\del_\tau\theta$ is
smaller towards the upper end of this range (as it should be clear by
inspection of the potential in Fig.~\ref{fig:theta}), therefore there is
a larger probability to find the meson in that region than towards the
bottom of the D7--brane.

Furthermore, the result for $\Delta E$ in \eqnum{period} implies
 \beq\label{deltabarQ}
 \Delta E(\Omega_n)\,\equiv\,\frac{\kappa^2_n}{2\Omega_n}\,\sim\,
 \frac{1}{\theta_1}\quad \Longrightarrow\quad \kappa^2_n \,\sim\,
 \Omega_n^{2/3}\,,\eeq
or $\bar Q_n\sim \bar \omega_n^{1/3}$. This is in agreement with our
previous discussion of \eqnum{smallk} in Sect.~\ref{sec:LL}~: there is
enough to make a rather small change $\Delta \bar Q_n \sim n^{1/3} R_0$
in the meson virtuality in order to jump from one mode to another at
fixed energy, whereas one needs a substantially larger increase in the
energy of the meson, namely $\Delta\bar\omega_n\sim R_0^3$, in order to
make that jump at fixed virtuality. The above result for $\bar Q_n$ is
moreover of the same order as the fluctuations $\delta\bar Q$ in the
virtuality of the flavor current due to its energy uncertainty, thus
justifying the energy averaging performed in our previous calculation of
the spectral weight, in \eqnum{ImPi}.

\bigskip
\section*{Acknowledgments}

We would like to thank Jorge Casalderrey--Solana for interesting
discussions and a careful reading of the manuscript. We acknowledge
discussions and correspondence on related issues with Alfonso Ballon
Bayona, Henrique Boschi-Filho, and Nelson Braga. The work of E.~I. is
supported in part by Agence Nationale de la Recherche via the programme
ANR-06-BLAN-0285-01. The work of A.H.~M. is supported in part by the US
Department of Energy.
\appendix

\section{WKB approach to light--like mesons}
\label{app:WKB}

In this appendix, we shall use the WKB approximation to rederive the
results in Sect.~\ref{sec:exact} on the spectrum and the wavefunctions of
the light--like mesons in the limit of a large quantum number $n\gg 1$.
We have already noticed in Sect.~\ref{sec:LL} that the WKB method
correctly reproduces the spectrum \eqref{Omegan} except for the constant
shift 1/6 in the eigenvalues $\Omega_n$ (which is of course a subdominant
term at large $n$). As we shall see, this constant shift is also the only
source of deviation between the WKB eigenfunctions and the asymptotic
expansion of the corresponding exact result, as given by \eqnum{psiLLas}.

Specifically, let us view \eqnum{ELL} as the Schr\"odinger equation for a
non--relativistic quantum mechanical particle in a stationary state with
energy $E=0$. We recall that the potential in that equation becomes
infinite at the end point $\z=1$ (cf. Fig.~\ref{fig:Vab} right). Then for
$\z < 1$, the associated wavefunction can be obtained in the WKB
approximation as \cite{LLQM}
 \beq\label{WKBpsi}
 \psi(\z)\,=\,\frac{C}{\sqrt{p(\z)}}\,\sin\left(
 \int_\z^1\rmd\z' \,p(\z')\,+\,\alpha\right)\,,
 \qquad p(\z)\,\equiv\,\sqrt{-V(\z)}\,.
 \eeq
The shift $\alpha$ in the argument of the sine function is not accurately
determined by the WKB method, but will be later fixed by matching onto
the exact solution near $\z=1$. With the potential $V(\z)$ in
\eqnum{ELL}, the integral in \eqnum{WKBpsi} is straightforward and yields
 %(up to an overall normalization)
 \beq\label{WKBpsi1}
 \psi(\z)\,=\,C'\left(\frac{\z}{1-\z}\right)^{1/4}\,\sin\left[
 2\Omega\left(\frac{\pi}{2} - \arcsin\sqrt{\z} - \sqrt{\z(1-\z)}
 \right)\,+\,\alpha
 \right]\,.
 \eeq
This solution is not reliable very close to the end points at $\z=0$
(where the potential becomes singular) and $\z=1$ (where the potential is
not differentiable), but it should be a reasonable approximation at the
intermediate points. A similar WKB solution can be constructed on the
right of $\z=0$. The condition that the two solutions match with each
other at intermediate points leads to the Bohr--Sommerfeld quantization
condition \eqref{BS}, which in turn implies $\Omega_n\approx n$ when
$n\gg 1$. With $\Omega_n=n$, the WKB solution \eqref{WKBpsi1} has exactly
$n$ nodes in between 0 and 1.

When approaching the end point at $\z=1$, we expect \eqnum{WKBpsi1} to
remain a good approximation so long as \cite{LLQM}
 \beq\label{WKBvalid}
 (1-\z)^{3/2}\,\gg\,\frac{1}{\sqrt{-\rmd V/\rmd\z}}\,=\,\frac{1}{2\Omega}
 \,,\eeq
where $-\rmd V/\rmd\z =4\Omega^2$ is the left derivative of the potential
at $\z=1$. This range of validity, which in terms of the variable $\xi$
introduced in \eqnum{xiOmega} amounts to $\xi\gg 1$, is wide enough to
allow for a matching between the WKB solution and the exact solution near
$\z=1$. The latter is the following linear combination of Airy functions
  \beq\label{psiAiry}
 \psi_{\rm b}(\xi)\,\simeq\,\frac{\sqrt{3}}{2}\,
 {\rm Ai}(-\xi)\,-\, \frac{1}{2}\,
 {\rm Bi}(-\xi)
 \qquad\mbox{for \ \ $0\,\le\, \xi\,\ll\,(2\Omega)^{{2}/{3}}$}
 \,,\eeq
where the relative coefficient has been fixed by the condition that
$\psi=0$ when $\xi=0$. Clearly, Eqs.~(\ref{WKBpsi1}) and \eqref{psiAiry}
have a common validity domain at $1\ll \xi \ll (2\Omega)^{{2}/{3}}$, and
within that domain they are indeed consistent with each other, as we now
check.

To that aim, we need to compare the approximate form of \eqnum{psiAiry}
valid at $\xi\gg 1$, which is obtained similarly to \eqnum{psiLLinter}
and reads
 \beq\label{psiAiryas}
 \psi_{\rm b}(\xi)\,\simeq\,
 \frac{C}{\xi^{1/4}}  \,\sin\left(\frac{2}{3}\,\xi^{3/2}\,
 + \frac{\pi}{12}\right),
 \eeq
to the approximate form of \eqnum{WKBpsi1} valid near $\z=1$ (\ie for
$\xi \ll (2\Omega)^{{2}/{3}}$), which is readily obtained as
  \beq\label{WKBpsi2}
 \psi(\xi)\,\simeq\,
 \frac{C}{\xi^{1/4}}
 \,\sin\left(\frac{2}{3}\,\xi^{3/2}\,+\,\alpha\right)
 %\quad\mbox{for}\quad 1\,\ll\,\xi\,\ll\,(2\Omega)^{{2}/{3}}
 \,.\eeq
Clearly, these two expressions, \eqref{psiAiryas} and \eqref{WKBpsi2},
are consistent with each other, as anticipated. Moreover they perfectly
match provided one choses $\alpha=\pi/12$, thus completely fixing the WKB
solution \eqref{WKBpsi1}, valid for any $\z$ which obeys the condition
\eqref{WKBvalid} and which is not too close to $\z=0$.

%\bibliographystyle{utcaps}
%\bibliography{ADSref}
%\end{document}

\providecommand{\href}[2]{#2}\begingroup\raggedright

\endgroup

\end{document}